\documentclass[twocolumn]{aastex63}

\usepackage{float}
\usepackage{graphicx}
\usepackage{amsmath}
\usepackage{hyperref}
\usepackage{natbib}
\newcommand{\farcsec}{\hbox{$.\!\!^{\prime\prime}$}}
\newcommand{\farcmin}{\hbox{$.\!\!^{\prime}$}}

\received{}
\revised{}
\accepted{}
\submitjournal{PASP}

\shorttitle{Science Commissioning of NIHTS}
\shortauthors{Gustafsson et al.}

\begin{document}

\title{Science Commissioning of NIHTS: The Near-infrared High Throughput Spectrograph on the Lowell Discovery Telescope}

\correspondingauthor{Annika Gustafsson}
\email{ag765@nau.edu}

\author[0000-0002-7600-4652]{Annika Gustafsson}
\affiliation{Department of Astronomy and Planetary Science, 
Northern Arizona University,
P.O. Box 6010, 
Flagstaff, AZ 86011}

\author[0000-0001-6765-6336]{Nicholas Moskovitz}
\affiliation{Lowell Observatory, 1400 W Mars Hill Rd., Flagstaff, AZ 86011}

\author{Michael C.\ Cushing}
\affiliation{Department of Physics and Astronomy, University of Toledo, 2801 W. Bancroft St., Toledo, OH 43606}

\author{Thomas A.\ Bida}
\affiliation{Lowell Observatory, 1400 W Mars Hill Rd., Flagstaff, AZ 86011}

\author{Edward W.\ Dunham}
\affiliation{Lowell Observatory, 1400 W Mars Hill Rd., Flagstaff, AZ 86011}

\author{Henry Roe}
\affiliation{Gemini Observatory/AURA, Santiago, Chile}

\begin{abstract}
The Near-Infrared High Throughput Spectrograph (NIHTS) is in operation on the 4.3 m Lowell Discovery Telescope (LDT) in Happy Jack, AZ. NIHTS is a low-resolution spectrograph (R$\sim$200) that operates from 0.86 to 2.45 microns. NIHTS is fed by a custom dichroic mirror which reflects near-infrared wavelengths to the spectrograph and transmits the visible to enable simultaneous imaging with the Large Monolithic Imager (LMI), an independent visible wavelength camera. The combination of premier tracking and acquisition capabilities of the LDT, a several arcminutes field of view on LMI, and high spectral throughput on NIHTS enables novel studies of a number of astrophysical and planetary objects including Kuiper Belt Objects, asteroids, comets, low mass stars, and exoplanet hosts stars. We present a summary of NIHTS operations, commissioning, data reduction procedures with two approaches for the correction of telluric absorption features, and an overview of select science cases that will be pursued by Lowell Observatory, Northern Arizona University, and LDT partners.
\end{abstract}



\section{Introduction} \label{sec:intro}

The Near-Infrared High Throughput Spectrograph (NIHTS, pronounced ``nights") is a low-resolution ($R\equiv \lambda / \Delta \lambda \sim 200$) near-infrared prism spectrograph on the 4.3 m Lowell Discovery Telescope (LDT), formerly the Discovery Channel Telescope, in Happy Jack, AZ (altitude 2347~m). NIHTS captures a wavelength range of 0.86-2.45~microns covering \emph{YJHK} bandpasses. The instrument contains no moving parts and employs a single, variable-width slit mask with eight different slit widths ranging from 0.27 to 4.03~arcseconds, each 12~arcseconds in length. The design, construction, and laboratory performance of NIHTS are discussed in detail in \cite{2018Dunham} and summarized in Section~\ref{sec:overview}. In this work, we report on instrument commissioning and the performance of NIHTS on sky at the LDT. We discuss observing procedures in Section~\ref{sec:obs}, instrument performance in Section~\ref{sec:performance}, data reduction techniques in Section~\ref{sec:reduction}, and several key science cases in Section~\ref{sec:science}.



\section{Instrument Concept and Design} \label{sec:overview}
NIHTS is the final of the first-generation instruments on the LDT and was designed specifically to target faint objects and allow for simultaneous visible imaging through the use of a custom dichroic. NIHTS is permanently mounted on the LDT instrument cube on a side port at the \emph{f}/6.1 Cassegrain focus of the telescope. The instrument cube \citep{2012Bida}, capable of accommodating five instruments, provides the necessary guiding, wave front sensing, and rotation for telescope operations. NIHTS is one of three instruments always on the instrument cube along with the visible Large Monolith Imager (LMI) and visible spectrograph DeVeny \citep{2014Bida}. The entire cube can rotate to compensate for the alt-az motion of the telescope, although a fixed rotator mode where the instruments remain fixed relative to the alt-az coordinate system is also possible.

\begin{figure}[!t]
    \centering  
    \includegraphics[width=\linewidth]{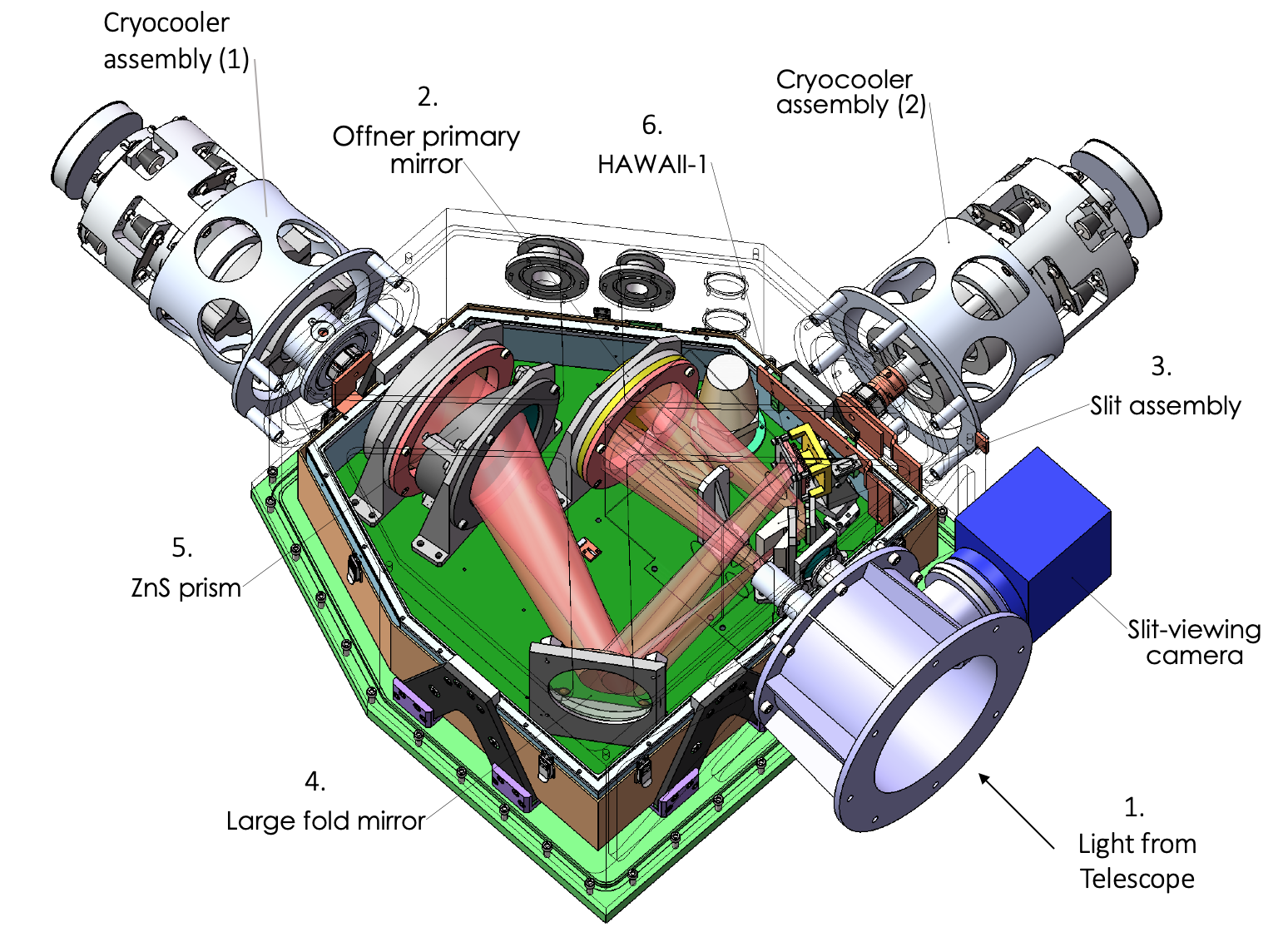}  
    \caption{NIHTS optical layout.} 
    \label{fig:NIHTS_Layout}
\end{figure}

The optical layout of NIHTS is compact with no moving parts; its longest dimension is just under 50~cm (see Figure~\ref{fig:NIHTS_Layout}). Light enters the telescope through the instrument cube (1), proceeds through an Offner relay (2), and is redirected to NIHTS via a deployable dichroic mirror that allows for simultaneous visible imaging using LMI. The light directed to NIHTS enters an anti-reflection coated dewar window and comes to a focus at the reflective slit plate (3). The instrument also contains a cold stop which is slightly oversized to ensure maximum throughput yet increases thermal background at longer wavelengths. The cold stop was tested by \cite{2018Dunham} and appears to reduce signal by about half at the edge of the LDT primary battle and reduces the signal all the way to zero at 100~mm outside the primary baffle's edge. Precision-cut baffles are located at focus, the Offner mirrors, enclosing the Offner relay assembly, and enclosing separately the slit optical section \citep{2014Bida}. The dichroic mirror was designed with a unique custom coating applied by Omega Corp. that reflects at high efficiency to the NIHTS detector from 0.9 to 2.5 microns (96.8\% average reflectance) while efficiently transmitting to LMI from 0.4 to 0.7 microns (90.1\% average transmission) \citep{2018Dunham}.

\begin{figure}[!t]
    \centering  
    \includegraphics[width=\linewidth]{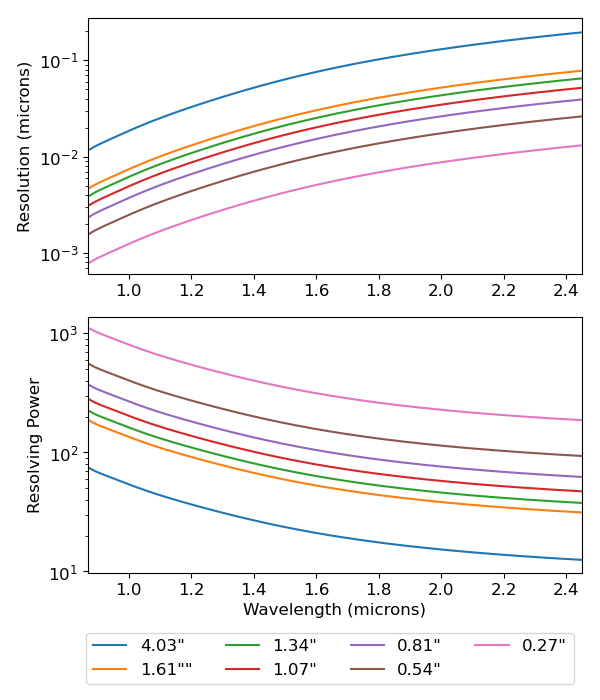}  
    \caption{Slit-limited resolution (top) and resolving power (bottom) for each slitlet measured across the full wavelength range. At short wavelengths, the resolving power range is 75-1100 and at long wavelengths the range is  15-200.} 
    \label{fig:resolution}
\end{figure}

The fixed slit mask has a single 96$''$ long slit divided into eight segments of different widths, referred to as slitlets, in a configuration with the widest slitlets near the edge of the mask, and the narrowest slitlets in the middle. Each slitlet is 12$''$ long and selectable by adjusting the telescope pointing. The eight slitlet widths in order across the mask are 4\farcsec03, 1\farcsec34, 0\farcsec81, 0\farcsec27, 0\farcsec54, 1\farcsec07, 1\farcsec61, and 4\farcsec03. The 0\farcsec27 slitlet (2~pixels) enables maximum resolution, and the two wide 4\farcsec03 slitlets (31~pixels) are included for flux calibrated spectral energy distribution (SED) modes in which no slit losses are expected under typical LDT conditions (median zenith seeing is 0\farcsec97 in \emph{V}-band; \citealt{2016Levine}). Most observing has been done in the 0\farcsec81 or 1\farcsec34 slitlets to best match seeing conditions. The slit-limited resolution and resolving power of each slitlet are shown in Figure~\ref{fig:resolution}. The mean resolving power across all wavelengths and slit widths is $\sim$200.

An advantage to the fixed slit plate is the ability to quickly offset between the SED-modes and the narrower slitlets. A typical observing sequence will involve numerous short (60-120~s) background-limited exposures, each dithered between nod positions on the slitlet of choice. For a given target, a subset of  exposures could be taken through the SED slitlet and used to flux calibrate the final, higher resolution spectrum.

Light reflected off the slit plate is refocused and exits the dewar to a slit-viewing camera with throughput from 0.9-1.4~microns. The light transmitted through the slit is redirected to the ZnS (Cleartran) prism (5) by a gold-coated fold mirror (4) and onto the NIHTS detector (6), mounted at 25.9$^{\circ}$ to the beam in order to meet focus over the full spectral range \citep{2014Bida}. The optical configuration of the double-pass prism and spherical back-mirror is based on the design of the Aerospace Corporation's Broadband Array Spectrograph System \citep{1997Warren} and provides an excellent instrumental PSF ($<<$1 pixel FWHM everywhere). 
 
The spectroscopic detector on NIHTS is a 256$\times$256 HgCDTe HAWAII-1 array manufactured by Rockwell Scientific and is fully utilized by the 96$''$ slit mask and 256 wavelength channels from 0.86 to 2.45~microns. The quantum efficiency (QE) is 50-64\% across the wavelength range of the detector. The entire assembly is chilled with two Sunpower closed-cycle Stirling cryocoolers. The compact size and lack of moving parts helps to create a rigid instrument and avoid data calibrations issues associated with flexure. At this time, exposure times for the smallest slit widths do not appear to be limited by the amount of thermal background in the widest slits. We do not see any evidence for persistence, but this has not yet been carefully measured and quantified.

\subsection{Slit-Viewing Camera and Guiding}
The slit-viewing camera is used for acquiring targets and confirming that the targets remain on the slit during spectral integrations. The slit-viewing camera is a thermo-electrically cooled Xeva-1.7-320 InGaAs array from Xenics that has high QE over 0.9-1.7~microns. A 1.37~micron shortwave pass filter is used to block the bright \emph{H}-band OH lines and maximize acquisition sensitivity in a wide \emph{Y} to \emph{J} bandpass ($\sim$0.9-1.4~microns). Under typical good observing conditions, we can locate a \emph{J}=16 target with 15 coadded 4~s exposures. 

Guiding is performed using the facility Guider and Wave front Sensing System (GWAVES) \citep{2012Venetiou}. GWAVES has two deployable sensor arms, each of which can be configured for high-frequency (5~Hz) guiding or low frequency (30~s cadence) wave front sensing on off-axis sources. This lower frequency wave front sensing is used to maintain focus and continually adjust the primary mirror supports to maintain optimum image quality. GWAVES is capable of guiding in a variety of modes: at sidereal rates, at constant non-sidereal rates, at time variable non-sidereal rates calculated for solar system objects using a JPL Horizons-generated ephemeris, with the instrument rotator fixed (e.g., at the parallactic angle), and across pointing dithers (e.g., when nodding the telescope to move sources along the NIHTS slit). We do not currently have the ability to guide on images from the slit-viewing camera or the visible pass-through to LMI, but may implement these features in the future. The GWAVES cameras have Johnson UBVRI filters available for guiding, but the most common filter used during guiding is \emph{R}-band to minimize any differential refraction between the NIHTS target and the guide star.

In typical operations, the slit-viewing camera is used to acquire the target and center it on the slit, while simultaneously the GWAVES probes can be set-up on their guide stars. During spectral observations, GWAVES can maintain telescope guiding, focus, and primary mirror figure. Meanwhile, a continual sequence of exposures are acquired on the slit-viewing camera to confirm the target remains on the slit and observing conditions remain stable. The ability to continuously monitor and correct pointing errors or mitigate telescope drift improves operational efficiency.   

\subsection{Software System Design}

Presently there are four fundamental software systems used for operating and monitoring NIHTS. Each system runs on a different server, though the observer typically interfaces with them through a single observing computer.

Lowell Observatory Instrumentation System (LOIS) is the instrument control software system that controls NIHTS and many other instruments built at Lowell Observatory, NASA Ames Research Center, MIT, and Boston University \citep{2004Taylor}. It is a base-level system that runs the CCD controller and generates the FITS headers.  

{\it xcam} is a Mac Mini accessed through VNC screen sharing and is the primary interface for controlling the NIHTS slit-viewing camera and NIHTS spectral channel. There are two primary interfaces on {\it xcam}: a GUI panel called {\it NIHTS} and a GUI panel called {\it ZTV}. Both have corresponding terminal windows that display status information. Primary instrument control is through this NIHTS GUI through which the observer issues commands to control exposures on the slit-viewing camera and the NIHTS spectral channel (e.g., defining the type of exposure, exposure time, setting the desired slitlet). The NIHTS GUI employs a python interface that has the ability to send commands to the telescope control system via a STOMP\footnote{STOMP is a Python client library for accessing messaging servers, including ActiveMQ (https://pypi.org/project/stomp.py/.)} message broker. The NIHTS GUI is written in PyQt5, a Python binding of the cross-platform GUI toolkit Qt developed by Riverbank Computing. The {\it ZTV} GUI is an astronomical image viewer originally designed by Henry Roe\footnote{https://github.com/henryroe/ztv} that has been adapted to display and analyze images from the NIHTS slit-viewing camera. The ZTV GUI is written using the Python TKinter package.

{\it LOUI}  (Lowell Observatory User Interface) acts as a status summary system for the user. Where ZTV and the NIHTS GUI are used for instrument control, we use the LOUI to monitor instrument and telescope subsystem processes. LOUI also provides real time image display of the NIHTS spectral channel, the ability to perform limited image analysis, and view a summary of facility status. However, different from other LDT instruments, the NIHTS LOUI is generally not used to take exposures.

Java Operator Executive (JOE) is a software ``controller" that is primarily concerned with anything on the telescope that moves. JOE communicates with the telescope control system as well as the various LOUI user interface programs and LOIS camera control programs. JOE facilitates Wave Front Sensor control which adjusts the shape of the LDT primary mirror. JOE also communicates with various pieces of software when the telescope is guiding. This communication is via the Apache ActiveMQ Broker.\footnote{http://activemq.apache.org} ActiveMQ is an open source broker message system written in Java and allows connectivity through many cross languages, including Python.  

\section{Observing Procedure} \label{sec:obs}

Calibrations for NIHTS include arc line spectra from a Xenon lamp mounted inside the instrument cube and dome flat field exposures using dome lamps. Batch scripts are provided in the NIHTS GUI to collect relevant calibration frames for each of the NIHTS slitlets. The batch scripts execute a series of exposures with specific exposure times which have been optimized for each slitlet to avoid saturation from the thermal background at the longest wavelengths. 

\begin{figure}[!b]
    \centering  
    \includegraphics[width=\linewidth]{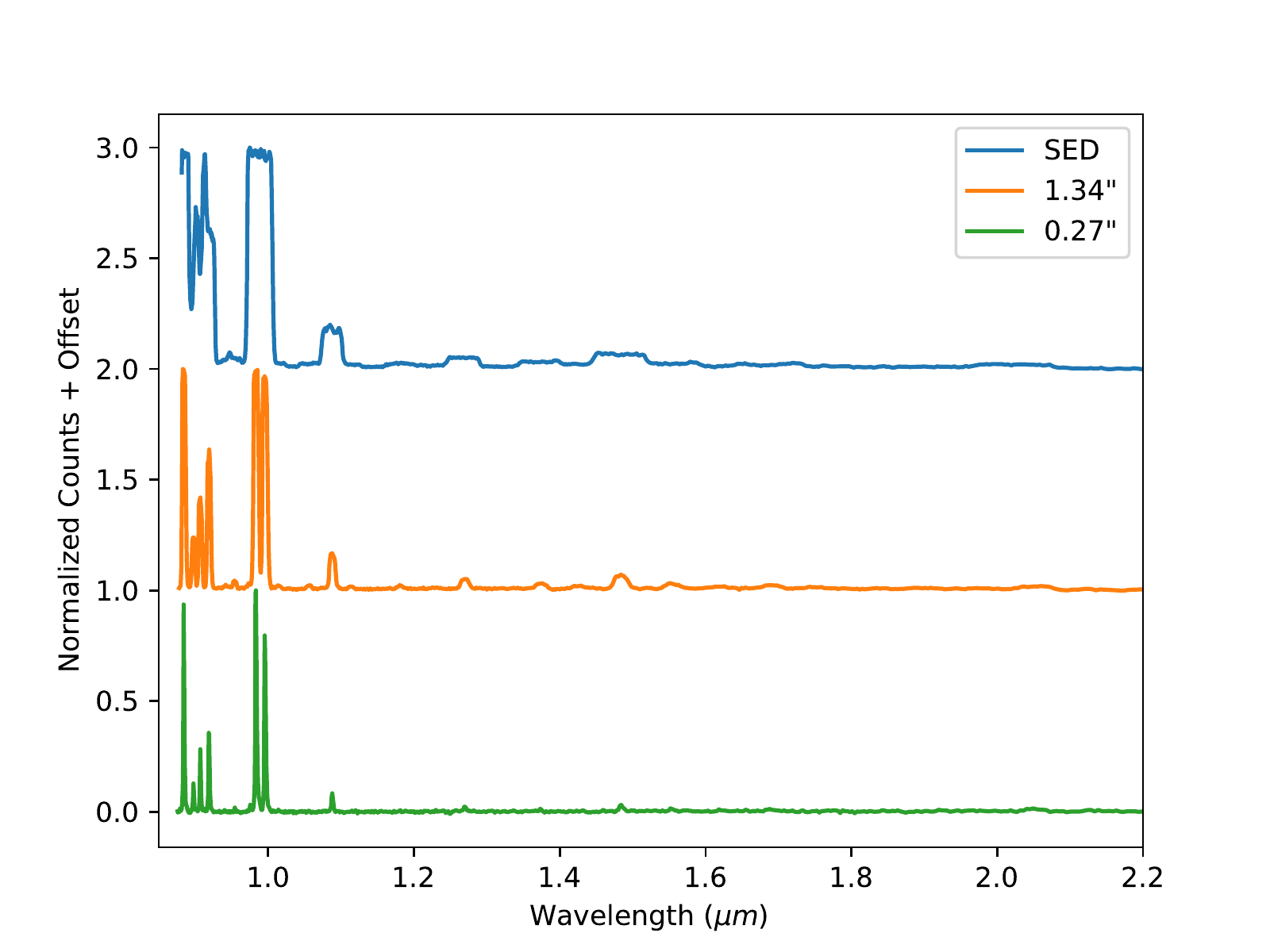}  
    \caption{Thermal background subtracted 120~s Xenon arc lamp exposures in three different slit widths: SED (4\farcsec03), 1\farcsec134, 0\farcsec27.} 
    \label{fig:arcs_allslits}
\end{figure}

The Xe arc lamp generates calibration lines from 0.86 to 2.0~microns. An example of a 120~s Xe arc lamp spectrum for three different slit widths is shown in Figure~\ref{fig:arcs_allslits} representing the range of resolving power between the widest and narrowest slits. Dome darks and flats are collected at a screen inside the dome using 4700~K 12~V lamps run at half voltage. For both arcs and dome flats, we collect dark exposures of equivalent exposure times with the lamps off to provide a means for removal of the thermal background that dominates at long wavelengths ($\lambda > 2.3$ microns).

Once on sky, the observer can focus the instrument with a star on the slit-viewing camera (0.9-1.4~microns) by applying small piston offsets of order $\sim$50-75~microns to the secondary mirror (M2). The focus is performed on the slit-viewing camera, not the science detector, but the difference has been confirmed to be of order 1 focus step (50 microns) on M2 under typical seeing conditions \citep{2018Dunham}. 

\begin{figure}[!t]
    \centering  
    \includegraphics[width=\linewidth]{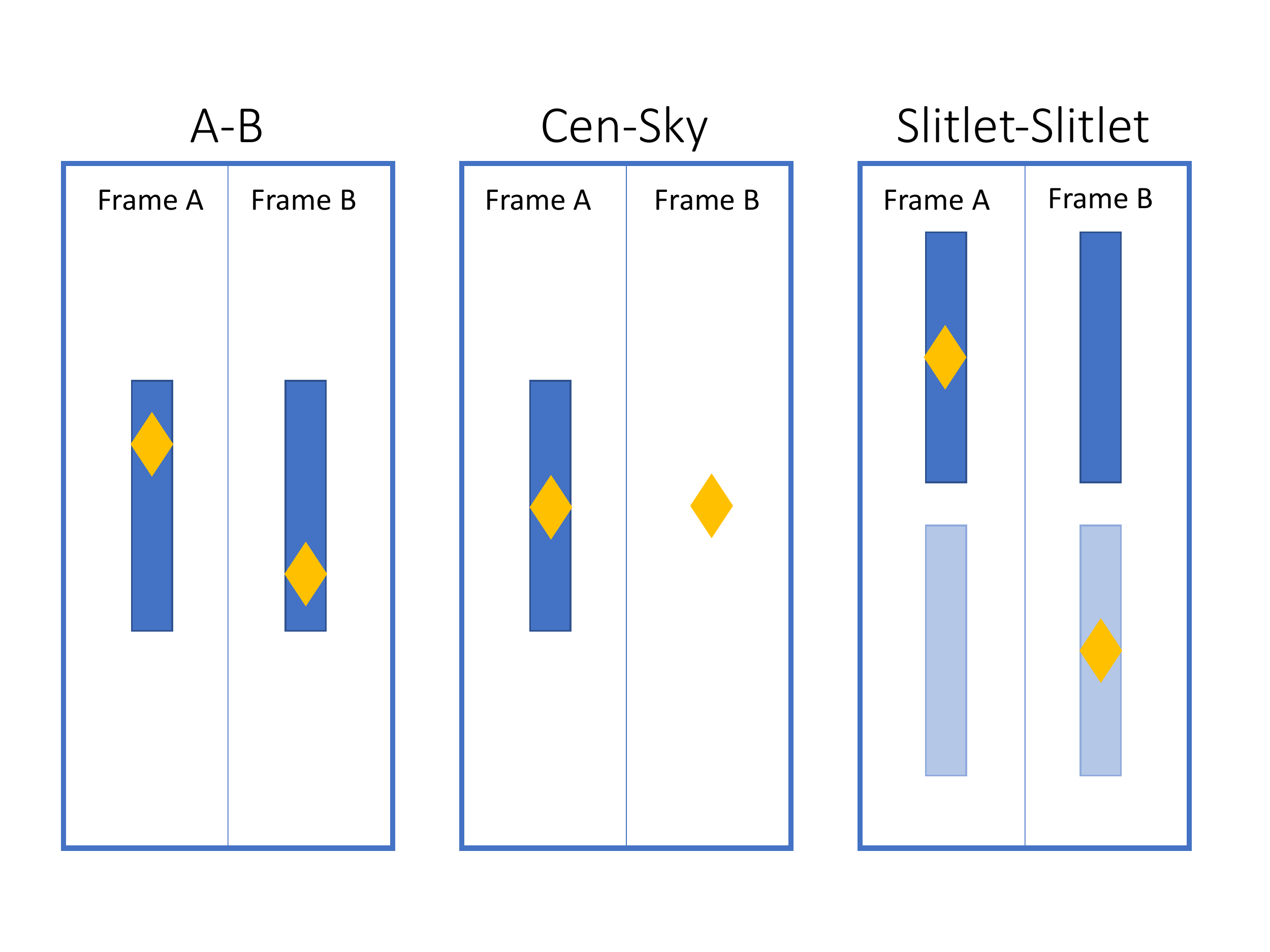}  
    \caption{The NIHTS GUI supports three different types of nod patterns. On Slit AB/ABBA: the target (diamond) is observed in both frame A and frame B at different positions along the slit separated by 5$''$. This ensures that both target and sky flux are collected during all exposures. Center-Sky: the target is observed in frame A and the sky is observed in frame B by offsetting the target from the slit in any user-specified direction (N, S, E, W, R.A., decl.) and distance ($''$). For this nod pattern, the target is only observed for half of the exposures. Slitlet to Slitlet: the target is first observed in slitlet 1 (dark blue) and then dithered to land on slitlet 2 (light blue) for frame B. (Figure modified from \citealt{2018Belli})} 
    \label{fig:Nods}
\end{figure}

Acquisition of a target onto a slitlet is achieved by placing the target at a designated {\it Home} location on the NIHTS slit-viewing camera. From {\it Home}, the observer can easily move the target with small telescope offsets to any predefined location on the focal plane (e.g., A-B nod positions for a given slitlet).

The NIHTS GUI currently supports collection of target spectra via single exposures of the spectral channel or by executing nod sequences including AB, ABBA, Center-Sky (on slit to off slit), and slitlet to slitlet nods between pairs of slitlets. See Figure~\ref{fig:Nods} for a visual representation of the nod sequences currently supported for NIHTS. The availability of all slitlets for each spectral exposure can also enable collection of simultaneous spectra of sources separated by less than the 96$''$ full length of the slit mask. For example, sources separated by $\sim$84$''$ can be simultaneously observed with the two SED (4\farcsec03) slitlets. An example science use case for the simultaneous observation of two targets separated by $84''$ is described in Section~\ref{subsec:exoplanets}. Other combinations of slitlets also enable simultaneous spectra, albeit at different resolution.

\subsection{Simultaneous LMI Observations}
\begin{figure}[!b]
    \centering 
    \includegraphics[width=\linewidth]{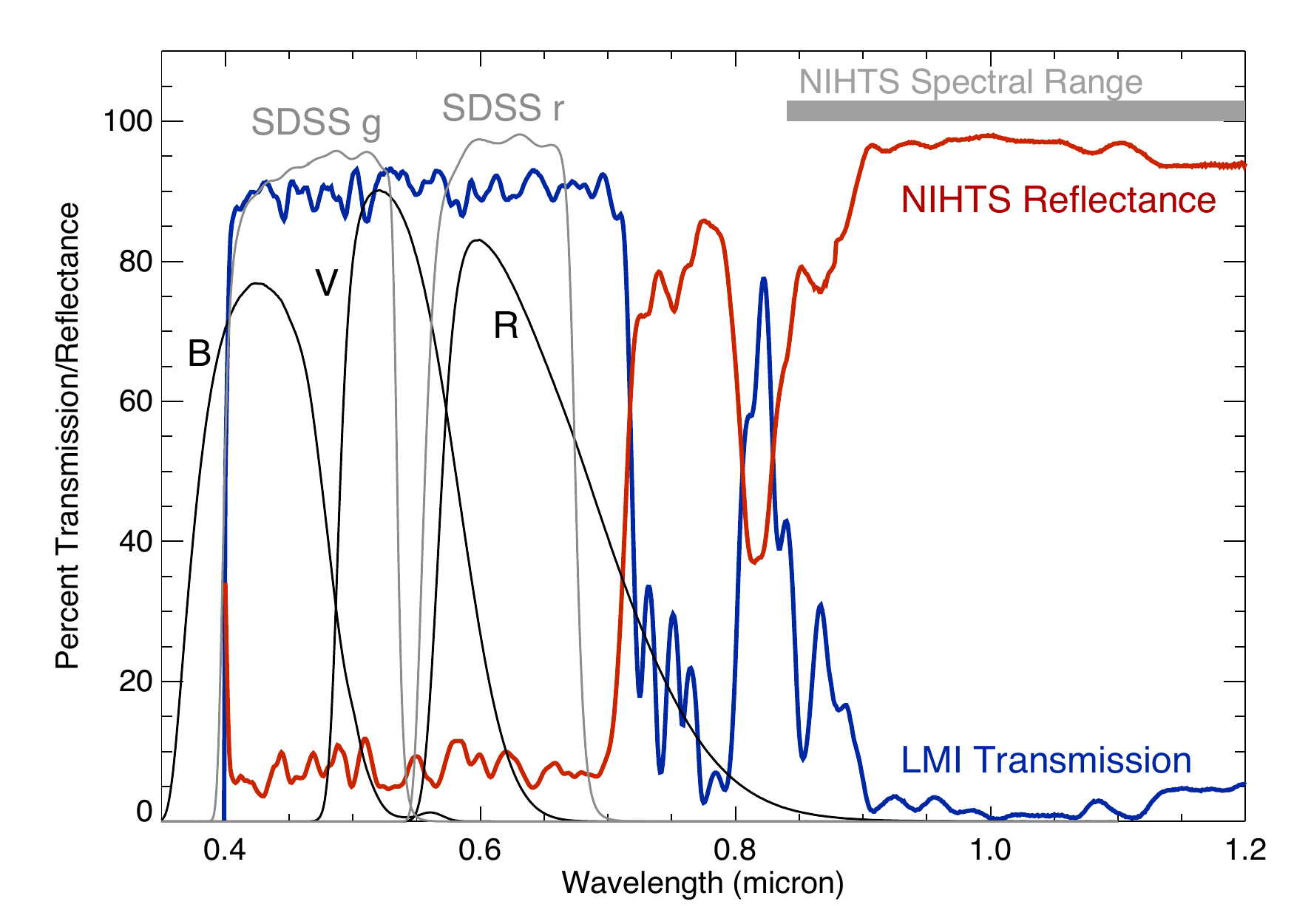}
    \caption{Dichroic transmission through to LMI and reflectance to NIHTS. The spectral range of NIHTS is shown. These curves are roughly constant longwards of 1.0 micron out to 2.4 microns. The transmission through to LMI spans wavelengths that overlap Johnson-Cousins \emph{BVR} and SDSS $g'$ and $r'$ filters.} 
    \label{fig:Dichroic}
\end{figure}

During early commissioning, NIHTS was fed by a fold mirror in the center of the instrument cube. This mirror was replaced by a dichroic on 13 December 2017 to allow for simultaneous LMI imaging and NIHTS spectroscopy. The dichroic transmits visible wavelengths to LMI and reflects the near-infrared (Figure~\ref{fig:Dichroic}). The visible wavelength transmission of the dichroic is about 90\% between 0.4 and 0.7~microns. An example of an LMI lightcurve of asteroid 93 Minerva obtained through the dichroic while collecting NIHTS spectra is shown in Section~\ref{subsec:Ast}. The dichroic is installed on a deployable and adjustable arm within the instrument cube. There is minimal flexure of the dichroic support arm. When the dichroic support arm is fully extended, maximum flexure is reached at $\sim 0\farcsec03$/15 degrees altitude.

The distance of the dichroic from the NIHTS instrument port can be adjusted in real time to optimize focus or to maintain focus as adjustments to the piston-offset of the secondary mirror are made (e.g., during LMI filter changes). The telescope control system is configured such that changing the LMI filter automatically causes a piston offset of the secondary mirror to accommodate for filter-dependent focus. The dichroic also then has to be offset to compensate for this change in light path to maintain focus on NIHTS. 

\begin{figure}[!t]
    \centering  
    \includegraphics[width=\linewidth]{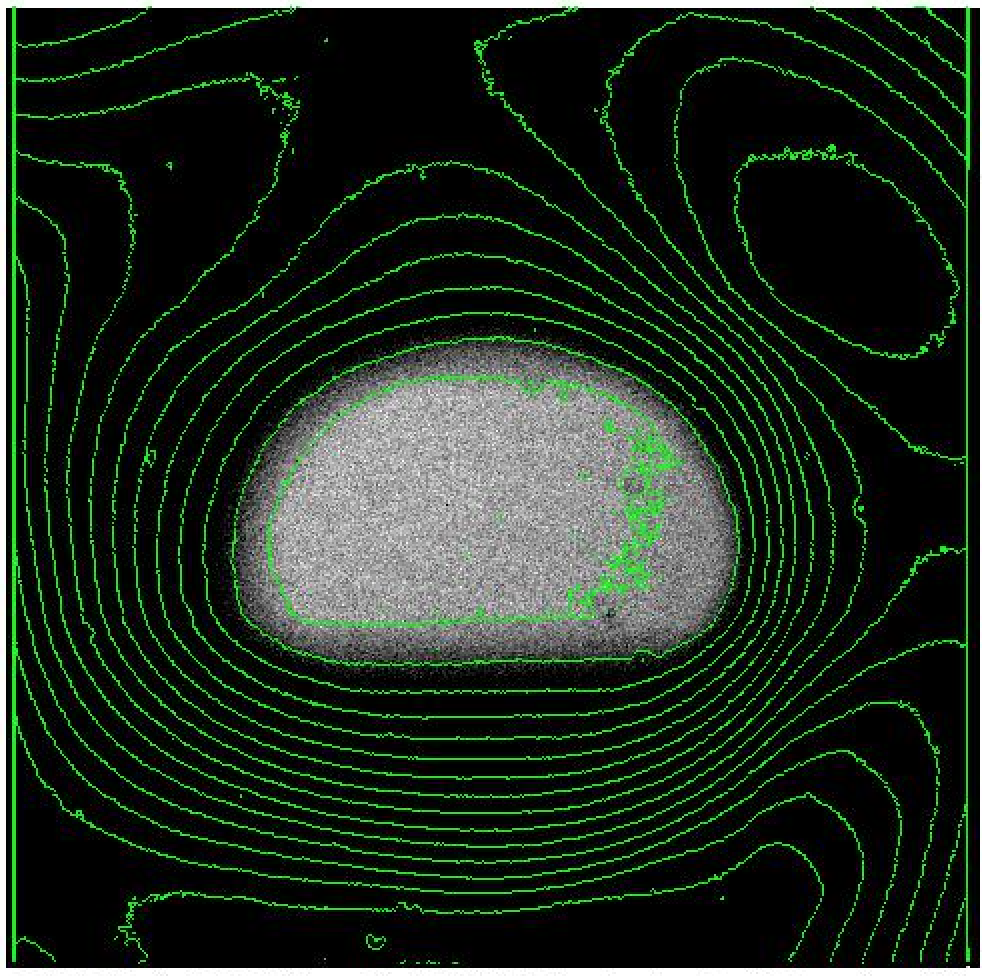} 
    \caption{Observable field of view on LMI with the NIHTS dichroic deployed. The unobstructed field is 6\farcmin4$\times$4\farcmin1, while the full unobstructed field of view shown as the full image frame is 12$\farcmin$3$\times$12$\farcmin$3. The green contours are at 500 ADU intervals between 15,000 and 25,000 counts. Image courtesy of Stephen Levine (Lowell).} 
    \label{fig:LMIFOV}
\end{figure}

Any changes to the dichroic position shift the LMI field of view.  The maximum vertical shift between LMI filters is less than 10$''$. The full LMI field of view (12$\farcmin$3$\times$12$\farcmin$3) is vignetted by the dichroic during simultaneous NIHTS$+$LMI observing. Figure~\ref{fig:LMIFOV} is an LMI flat field image taken in the SDSS~$r'$ filter with the NIHTS dichroic in the light path. The central region in gray is the unobstructed region roughly 6$\times$4~arcmin in size and throughput drops off steeply outside of that region.

\subsection{LMI Acquisition}
For acquisition exposures on the slit-viewing camera longer than a few seconds, thermal background saturates the unfiltered Xenics camera. Acquisition of targets down to magnitudes of $J\sim$16 can be achieved by co-adding short exposures from the slit-viewing camera. Objects fainter than this limit require a prohibitively large number of co-adds such that acquisition through the dichroic on LMI is instead recommended. This acquisition process requires mapping of LMI pixels to the NIHTS slitlet positions. Since flexure of LMI relative to NIHTS is not actively corrected, this pixel mapping is performed for each acquisition sequence and is achieved by imaging a bright field star ($V\sim$5--10) nearby the science target. Proximity of this field star to the science target is important to mitigate effects of atmospheric dispersion. The telescope is offset to position this field star at the desired location on the NIHTS slit mask, and the LMI pixel location of that star is noted. Finally, positioning the faint science target on that LMI pixel location effectively achieves a blind acquisition where the source may be easily seen on LMI but is lost in the high background of the NIHTS slit-viewing camera images. Visible sources much fainter (e.g., $H$=21.5; $V$=23) than the NIHTS limiting magnitude are easily detectable on LMI with exposures $\leq60$~s. Low mass stars and brown dwarfs near the NIHTS limiting magnitude that have very red colors can only be observed with LMI for blind acquisition if their visible colors are brighter than $\sim V$=24. The typical overhead required to acquire a target with LMI onto the NIHTS slit is $\sim$10-15 minutes. No further flexure between LMI and NIHTS has been noted, however, we recommend observers reacquire their target with changes in airmass to ensure the target remains on the NIHTS slit. 


\section{Instrument Performance} \label{sec:performance}

Following commissioning in early 2018, NIHTS has been available for normal science observing since 2018 July 1. In the first two years, NIHTS was requested for an average of 20 nights/semester for a variety of science cases, some of which are discussed in Section \ref{sec:science}.


\subsection{Detector Linearity and Saturation}

\begin{figure}[!t]
    \centering 
    \includegraphics[width=\linewidth]{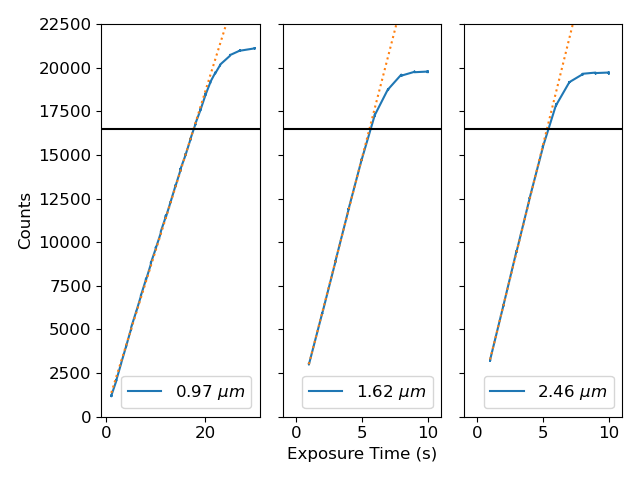}
    \caption{Linearity test with increasing exposure time vs counts at 0.97, 1.62, and 2.45~microns. We find the saturation limit at 16,000 counts.} 
    \label{fig:linearity}
\end{figure}

To first order, the NIHTS detector is linear up to about 16,000 counts at all wavelengths (Figure~\ref{fig:linearity}). The detector saturates faster at longer wavelengths due to added flux from the thermal background. Table~\ref{tab:saturation} shows the maximum wavelength that remains unsaturated in each of the slitlets with respect to a given exposure time. The NIHTS detector is sensitive up to 2.45~microns. For individual exposure times longer than about 45~s, it is difficult to obtain unsaturated data beyond 2.1~microns with any slitlet. Data have not been collected to probe saturation wavelength as a function of ambient temperature.

\begin{table*}
\caption{\label{tab:saturation} Maximum accessible wavelength (in microns) for a given slitlet and exposure time. NIHTS is sensitive to a maximum wavelength of 2.45~microns. The wider slitlets saturate in the longer wavelengths at much shorter exposure times than the smaller slitlets as a result of thermal background.}
\begin{tabular}{c c c c c c c c c } 
\hline \hline
Slitlet ($''$) & SED 1 	& 1.34 	& 0.81 	& 0.27  & 0.54  & 1.07  & 1.61  & SED 2 \\
\hline
ExpTime (s)     &       &       &       &       &       &       &       &    \\
 3              & 2.45  & 2.45  & 2.45  & 2.45  & 2.45  & 2.45  & 2.45  & 2.45 \\
 4              & 2.40  & 2.45  & 2.45  & 2.45  & 2.45  & 2.45  & 2.45  & 2.40 \\
 5              & 2.35  & 2.45  & 2.45  & 2.45  & 2.45  & 2.45  & 2.45  & 2.35 \\
 10             & 2.30  & 2.45  & 2.45  & 2.45  & 2.45  & 2.45  & 2.40  & 2.30 \\
 20             & 2.20  & 2.30  & 2.45  & 2.45  & 2.45  & 2.40  & 2.30  & 2.20 \\
 30             & 2.20  & 2.30  & 2.35  & 2.45  & 2.35  & 2.30  & 2.25  & 2.20  \\
 45             & 2.15  & 2.25  & 2.30  & 2.30  & 2.30  & 2.30  & 2.20  & 2.15 \\ 
\hline
\end{tabular}
\end{table*}


\subsection{NIHTS Efficiency}

\begin{figure}[!b]
    \centering 
    \includegraphics[width=\linewidth]{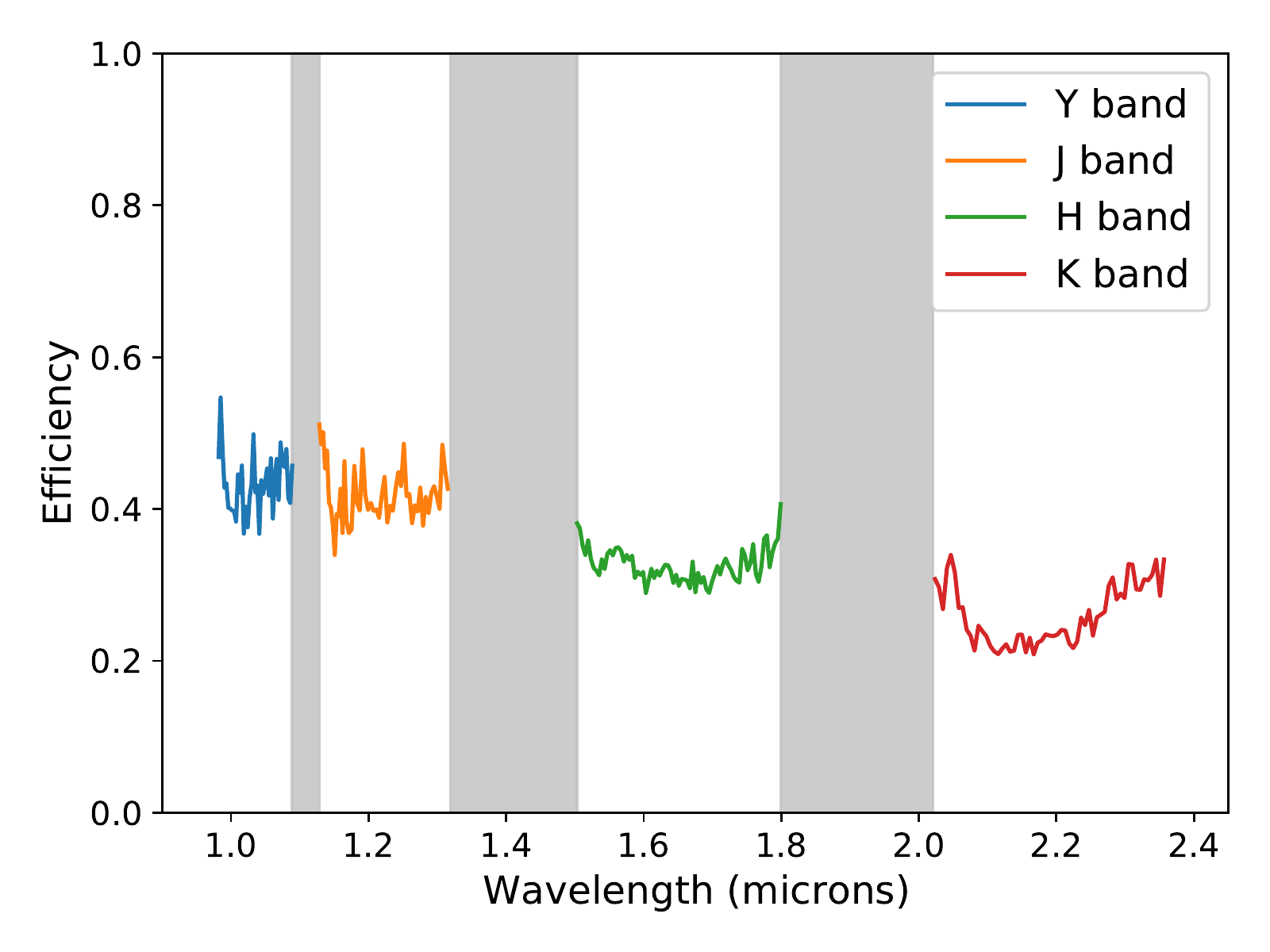}
    \caption{Measured NIHTS efficiency which accounts for  telescope, dichroic, and detector throughput in Y, J, H, and K bands. Regions of telluric absorption, which can have highly variable transmission, have been excluded (gray).}
    \label{fig:efficiency}
\end{figure}

We determine the efficiency of the NIHTS detector using spectrophotometric standard star Feige~110, observed through the SED 1 (4\farcsec03) slitlet. Weather conditions were clear with subarcsecond seeing ($\sim$0\farcsec9) and Feige~110 was at an airmass of 1.7. We compare the measured photon density (phot/s) of Feige~110 observed with NIHTS to the expected photon density of Feige~110 at the top of the atmosphere (Equation~\ref{eq:ratio}) using the AB magnitudes in \emph{Y, J, H}, and \emph{K} from the STSci CALSPEC database of STISNIC Hubble Space Telescope spectroscopy \citep{2014Bohlin}. We use the ratio of measured to expected photon density to derive the efficiency at each wavelength across the detector (Figure~\ref{fig:efficiency}).

\begin{equation} \label{eq:ratio}
    E = \frac{N_{\lambda,meas}}{N_{\lambda,exp}}
\end{equation}

We observed Feige~110 with 5~s NIHTS observations in the SED slitlet. NIHTS has a measured gain ($G$) of 24~e$^-$/ADU, read noise (RN) of 100$e^{-}$/pixel, and a mean QE of 60\%. Using this information and the $\Delta \lambda$ defined per pixel, we find the measured photon density at each wavelength ($N_{\lambda,meas}$) across the detector (Equation~\ref{eq:N_lam_m}) where $F_*$ is the flux from the star (ADU/s). 

\begin{equation} \label{eq:N_lam_m}
    N_{\lambda,meas} = \frac{F_* \times G}{QE(\Delta \lambda)}
\end{equation}

To determine the expected photon density incident upon the LDT aperture, we use calibrated near-infrared spectra of Feige~110 provided as $F_\lambda$. We convert from $F_\lambda$ to AB magnitude ($m_{AB}$) following \cite{1975Hayes}. Using the AB magnitude, we compute the expected number of photons ($N_{\lambda}$) for Feige~110 observed through a 1~m telescope as defined in \cite{1988Massey}:
\begin{equation} \label{eq:N_lam_exp}
    N_{\lambda, exp} = \frac{4.5\times10^{10}}{\lambda} 10^{-[(m_{AB} + A_{\lambda} \chi)/2.5]}.
\end{equation}

Here, $A_\lambda \chi$ is the extinction in magnitudes per airmass. Since we observed Feige~110 at an airmass of 1.7, we correct the spectrum to an airmass of 1.0 using a single conservative extinction coefficient of 0.06 for $A_\lambda$ across the entire wavelength range. We scale $N_{\lambda,exp}$ to the size of the LDT (4.3~m primary with a 1.4~m obscuration due to the secondary) and take 80\% of our final value to account for 1 airmass of atmospheric extinction \citep{1979Manduca}. 
Finally, we solve for the efficiency of NIHTS, including the LDT and instrument cube optics ($E$) using the ratio of the measured photon density to the expected photon density (Equation~\ref{eq:ratio}). 

We find an average efficiency across all wavelengths of approximately 38\%. If we instead adopt $A_\lambda$ of 0.03 following \cite{1975Hayes}, we find an upper limit on mean efficiency of $\sim$40\% thus bounding our predicted NIHTS efficiency between 38\% and 40\% across the full wavelength range.

\begin{figure}[!b]
    \centering 
    \includegraphics[width=\linewidth]{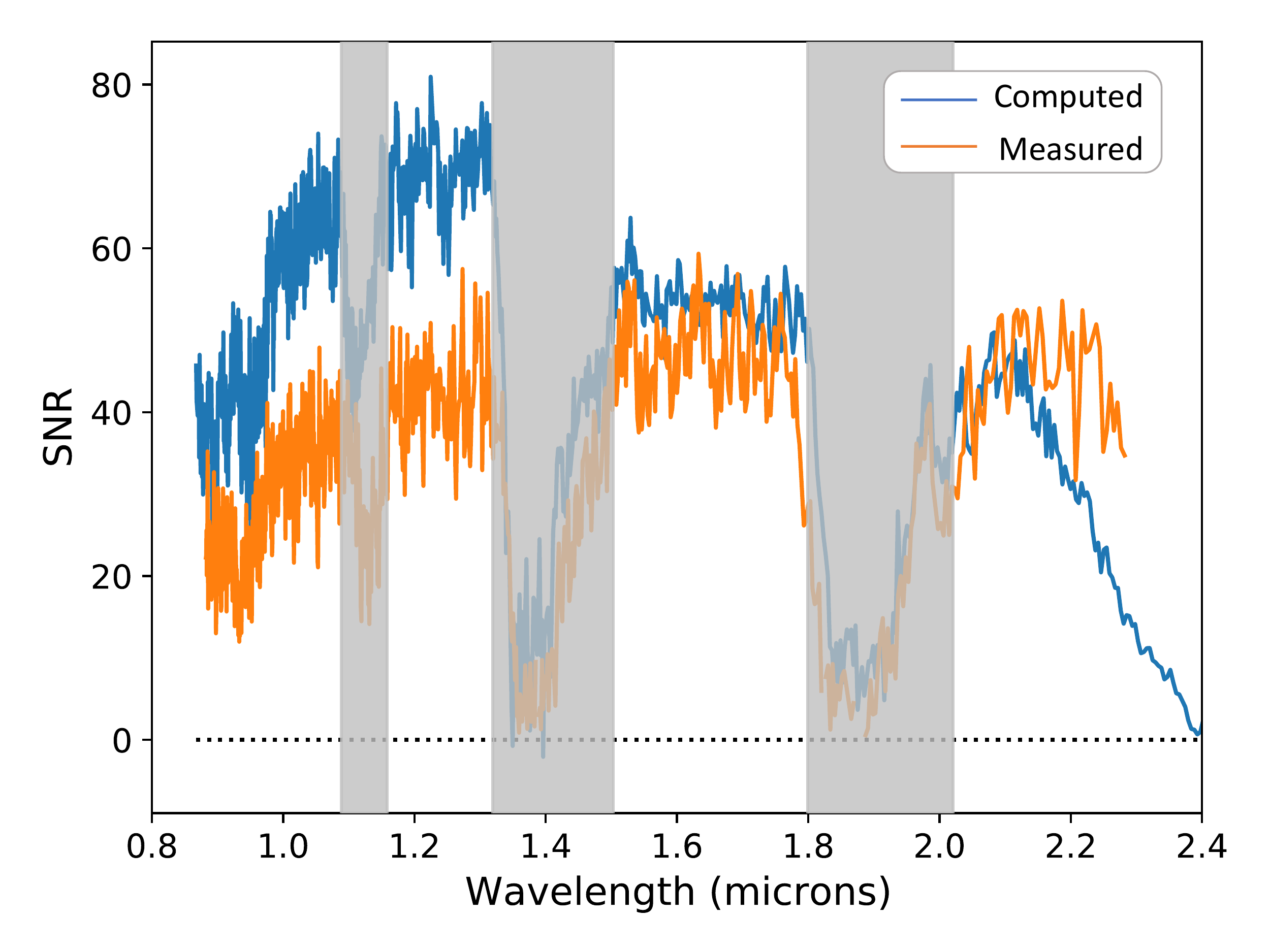}
    \caption{Computed vs. measured S/N for Feige 110 observations across the full wavelength range. We find the measured S/N to be lower in \emph{Y} and \emph{J} bands, but comparable in \emph{H} and \emph{K}.}
    \label{fig:SNR}
\end{figure}

We compute the signal to noise ratio (S/N) for each band as
\begin{multline} \label{eq:SNR}
   S/N = (R_* \times t)*[(R_* \times t) + (R_{sky} \times t \times n) \\ + (RN^2 + (\frac{G}{2})^2 \times n)]^{-1/2}
\end{multline}

where $R_*$ is the count rate from the star ($e^-$/sec), $R_{sky}$ is the count rate from the background ($e^-$/sec/pixel), $t$ is the exposure time (s), $n$ is the number of pixels in the aperture, $G$ is the gain (e$^{-1}$/ADU), and RN is the read noise ($e^-$/pixel)\footnote{\url{http://www.ucolick.org/~bolte/AY257/s_n.pdf}}. 

\begin{table*}
\caption{\label{tab:sensitivity_std} Measured counts integrated over different band passes for spectrophotometric standard star Feige 110. The S/N is provided as S/N per second of exposure time. The resulting mean efficiency of the instrument and telescope is given in the second to last column. The final column is the limiting magnitude calculated for each band for a S/N of 1 per 1-second of integration time.}
\begin{tabular}{c c c c c c c c c c} 
\hline \hline
 	    & 		        &  	       	&  	          &\multicolumn{3}{c}{Integrated Counts (4$''\times$12$''$ aperture)} 		&  		    &  \\ 
\cline{5-7}
Band    & $\lambda$     & Mag. 	    & Catalog Flux              &\# pixels   &Object   	    &Sky         	&S/N        & Mean Eff.  & Limiting Mag.\\
        & ($\mu m$)     &           &($erg/s/cm^2/\mbox{\AA}$)  &  	        & ($s^{-1}$)	&($s^{-1}$)     & ($s^{-1}$) 	        &(\%) & \\
\hline
Y       & 0.96--1.1    & 12.94     & 1.32$\times10^{-12}$      &2410       & 21550         & 20350         &445        & 43 & 19.8\\
J       &   1.1--1.4   & 12.55     & 2.54$\times10^{-13}$      &2090       & 22050         & 31500         &420        & 42 & 19.3\\
H       &   1.5--1.8   &  12.63    & 7.33$\times10^{-14}$      &1140       & 16800         & 76700         &260        & 33 & 18.7\\   
K       &   2.0--2.4   & 12.77     & 2.18$\times10^{-14}$      &860       & 10600         & 355400        &85        & 26 & 17.6\\
\hline
\end{tabular}
\end{table*}

The results for the total flux measurements and S/N are outlined in Table~\ref{tab:sensitivity_std}.  In the last column of the table, we derive the limiting magnitude integrated across each band for a S/N=1 at 1 second exposure time. We also compare the computed S/N across the full wavelength range for the measured S/N for Feige 110 observations. We find that the instrument under-performs in Y and J band, but is a good match in \emph{H} and \emph{K}.


\subsection{Dispersion Stability} \label{sec:disp_stability}

\begin{figure}[!t]
    \centering
    \includegraphics[width=\linewidth]{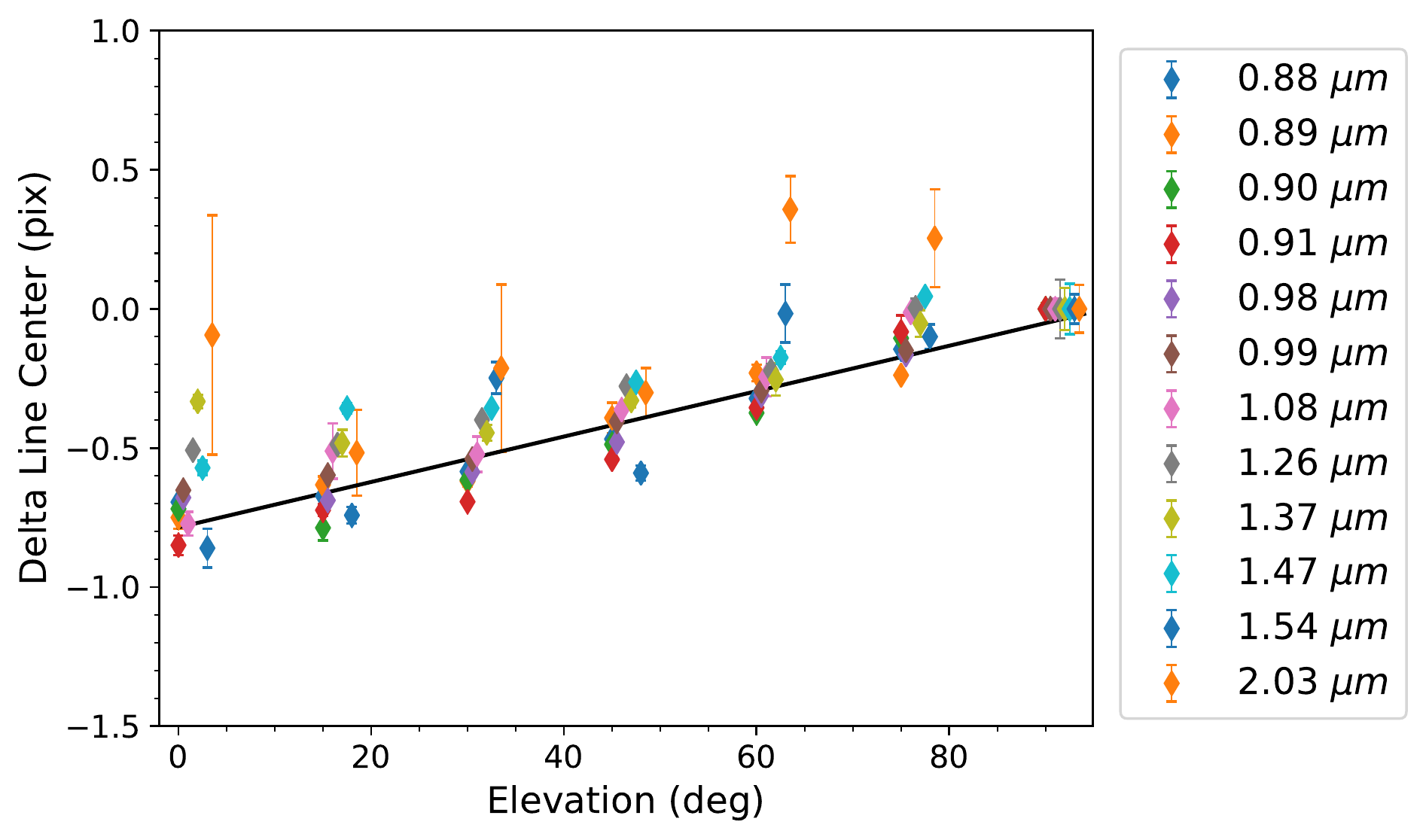}  
    \caption{Pixel shift of specific Xenon arc lamp lines as a function of telescope elevation. We sampled lines at wavelengths from 0.88 up to 2.03~microns with elevation changes in increments of 15$^{\circ}$ from 0$^{\circ}$ to 90$^{\circ}$. We find no more than a $\sim$1~pixel shift across the entire elevation range. We see the most scatter at the longest wavelength (2.03 $\mu m$) plotted in orange, which is attributed to the low signal-to-noise of that line. \label{fig:dispersion}}
\end{figure}

We quantify the dispersion stability in \emph{Y, J}, and \emph{H} bands up to 2~microns to assess flexure in the instrument. We collected 120~s arc lamp exposures at telescope pointings from 0$^{\circ}$ elevation to 90$^{\circ}$ elevation in increments of 15$^{\circ}$. At each pointing, we extract a single row in the center of the 1$\farcsec$34 slitlet and fit Gaussian profiles to the arc emission lines from 0.88~microns to 2.03~microns to identify the line centers. Figure~\ref{fig:dispersion} shows the fitted line centers with reference to the positions at 90$^{\circ}$ elevation. We find stability of the dispersion solution to less than 1 pixel at all elevation angles. The highest variation in arc line centers is seen at the longest wavelengths. However these lines also have the lowest signal to noise. The generally consistent shift of all arc lines across the spectrum as a function of elevation angle, suggests that arc lamp sequences can be taken at each target pointing and that just the brightest (highest signal-to-noise) arc lines can be used to account for any dispersion offset relative to the master dispersion solution. 

\section{Data Reduction} \label{sec:reduction}

Full spectroscopic datasets include calibration files (i.e., flat fields, arcs, and the corresponding dark frames), the science target, and standard stars (e.g., A0~V, G2~V). The key reduction steps for most spectral datasets include the creation of a normalized flat field image and wavelength calibration files, pair subtraction of A and B images, identification of the aperture positions, background subtraction and spectral extraction, and telluric correction. In the near-infrared, atmospheric absorption between the \emph{J, H} and \emph{K} windows is prominent in ground-based spectra (see Figure~\ref{fig:atm_trans}). Typical corrections for telluric absorption rely on observations of standard stars observed close in both time and airmass to the science target. This method is implemented in our standard data reduction procedure with Spextool (Section \ref{sec:spextool}). We also present an alternative technique for telluric correction of low-resolution NIHTS spectra using NASA Goddard's Planetary Spectrum Generator (PSG) tool (Section \ref{sec:psg}). For spectrophotometric calibration of the target, observations of a spectrophotometric standard must also be collected.

\subsection{Spextool} \label{sec:spextool}
 We have modified the Spextool package \citep{2004Cushing} to handle NIHTS reductions.  Spextool was originally written to reduce data obtained with SpeX \citep{2003Rayner}, a moderate resolution infrared spectrograph on the NASA Infrared Telescope Facility (IRTF). It has since been modified to support other instruments, most recently with significant changes to accommodate data from iSHELL \citep{2016Rayner}, the IRTF's new high resolution ($R\sim$70,000) infrared spectrograph.  The primary modification for iSHELL was to allow for the extraction of spectra when the dispersion and spatial axes of the spectral images did not align with rows and columns on the detector. The NIHTS version of Spextool takes advantage of these modifications as the spatial dimension of the slit image is tilted by a few pixels with respect to the array columns.  Xenon arc images taken at the beginning of the night are used for wavelength calibration purposes and to measure this distortion due to the tilt.
 
 \begin{figure}[!t]
    \centering
    \includegraphics[width=\linewidth]{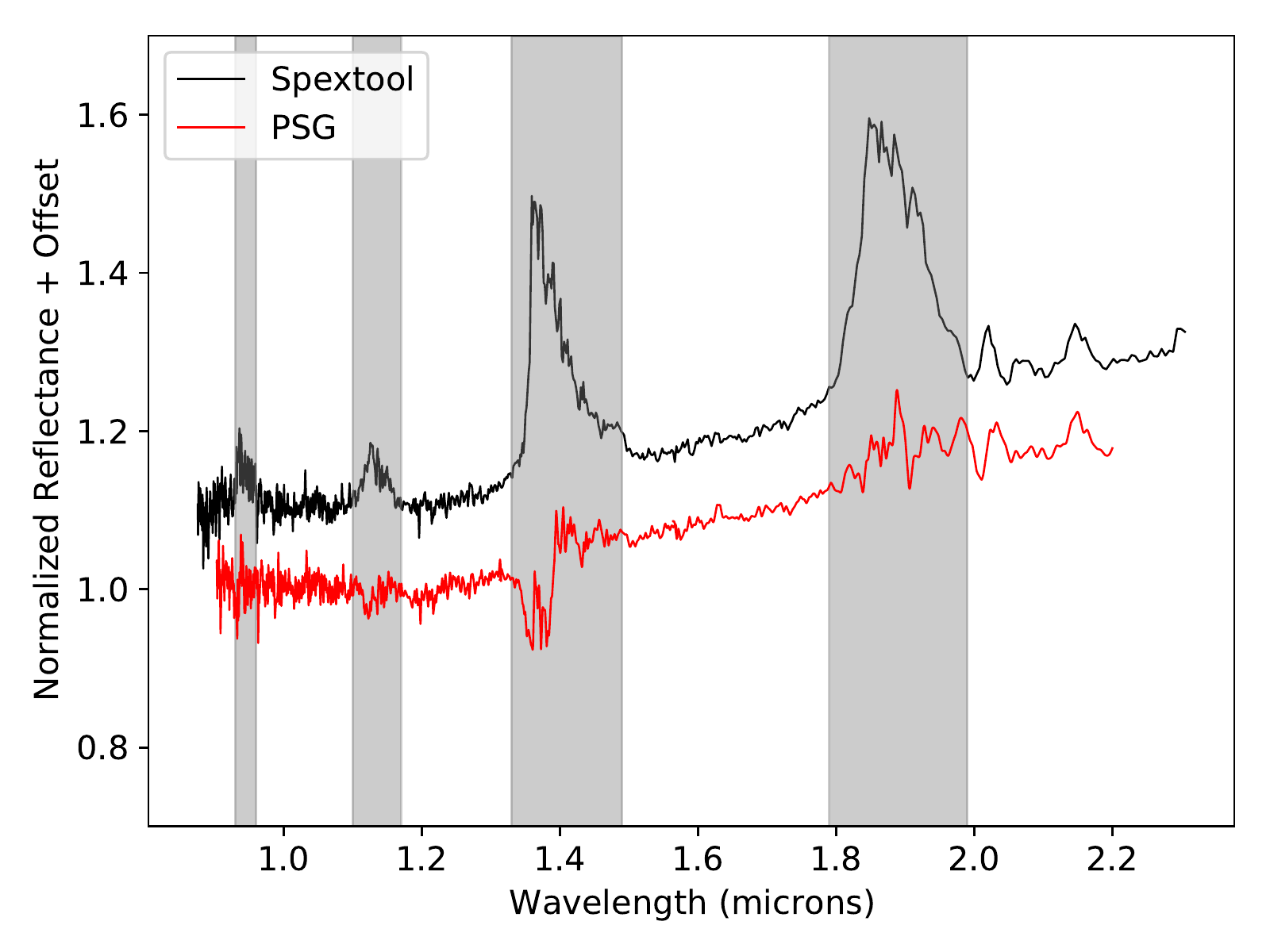}
    \caption{We compare a NIHTS reduction of Main Belt asteroid 93~Minerva with Spextool and PSG for observations collected under conditions with a high  precipitable water value (PWV). The PWV for these data was 5.26~mm , below the LDT annual mean PWV. The gray vertical bars represent regions with high telluric absorption. In the PSG reduction, we are able to correct most of those features leaving reduced residuals in comparison with the Spextool reduction. Full analysis of the NIHTS spectra of Minerva are presented in McAdam \& Gustafsson 2020 (submitted).}  \label{fig:Spextool_vs_PSG}
\end{figure}

Several additional modifications to the software were required in order to reduce NIHTS data, the most important of which was a modification to the telluric correction and flux calibration routines.  Spextool uses a model of Vega and observations of an A0 V standard star taken near in time and position to the science target in order to correct for telluric absorption \citep{2003Vacca}.  The Vega model undergoes several modifications before it is used to remove the intrinsic stellar spectrum from the standard star spectrum including smoothing to the average resolving power.  However, the resolution of NIHTS changes by a factor of greater than 15 from 1 to 2~microns, thus the smoothing routine was modified to allow for fast variable-resolution smoothing. 

A Spextool reduction of NIHTS data collected under highly variable conditions and poor seeing is shown in Figure~\ref{fig:Spextool_vs_PSG} of Main Belt asteroid 93~Minerva (McAdam \& Gustafsson, submitted). Minerva is a primitive C-type asteroid with a slightly red sloped spectrum and no major absorption or emission features \citep{1995Xu, 2002Binzel}. The large residuals centered around 0.9, 1.1, 1.4, and 1.9 microns cannot be further improved using the basic Spextool telluric correction routines due to variability and near saturation of the telluric bands (i.e., zero transmission through the atmosphere). Typically, when conditions are this poor, an atmospheric transmission (ATRAN) model can be adopted to potentially reduced the magnitude of residual telluric features (e.g., \citealt{2015Rivkin, 2019DeMeo}). In the following, we present an alternative approach that leverages PSG models to correct data when conditions are poor.


\subsection{Using the Planetary Spectrum Generator for Telluric Correction} \label{sec:psg}
\begin{figure}[!b]
    \centering
    \includegraphics[width=\linewidth]{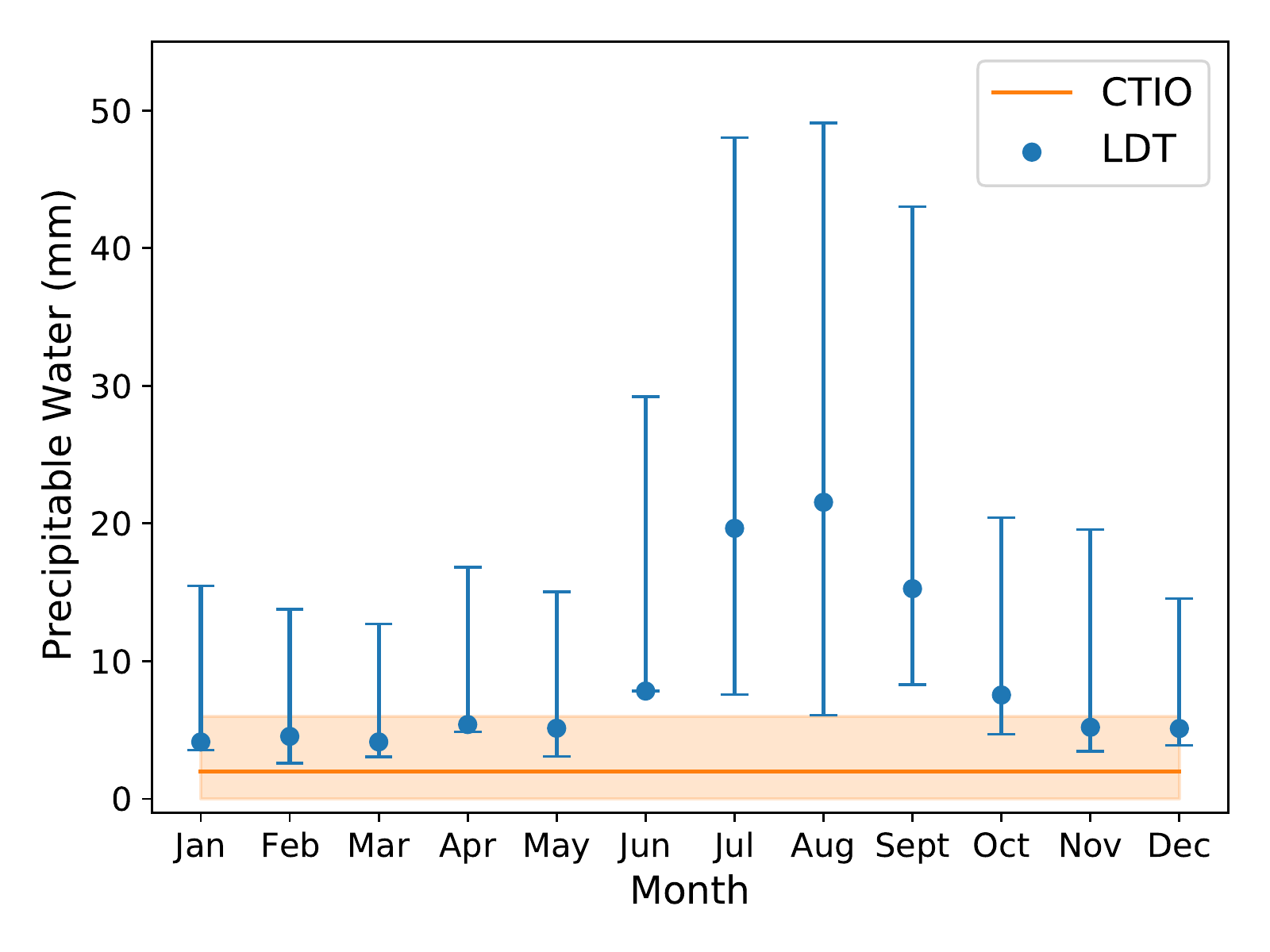} 
    \caption{The average and monthly range of precipitable water (mm) values for LDT from 2012 to 2013, against a 1-year average and range for CTIO over the same time frame \citep{2014Li} (orange). The LDT has a mean annual precipitable water of $8.8\pm 6.0$~mm and CTIO has an annual median precipitable water value of $2_{-2}^{+4}$~mm.} 
    \label{fig:DCT_PW}
\end{figure}

The strength of telluric absorption features in the near-infrared can be variable on timescales of just a few seconds and are highly sensitive to the level of precipitable water in the column of atmosphere above the observing site. As a way to quantify the relative atmospheric conditions at LDT, we compare the total precipitable water in a monthly average for LDT to an annual average for the Cerro-Tololo Inter-American Observatory (CTIO) which sits at an elevation of 2200~meters, only $\sim$150~m lower than the LDT (Figure~\ref{fig:DCT_PW}). \cite{2014Li} reports that CTIO has an annual precipitable water value (PWV) of 2~mm varying from 0~mm to 6~mm based on observations collected from 2012 to 2013. 

The LDT monthly average is obtained from the University of Wyoming upper air balloon sounding data\footnote{\url{http://weather.uwyo.edu/upperair/sounding.html}} which provides two measurements per day of the PWV (UT 0, UT 12) for Flagstaff (Station \#72376). We determine the monthly mean PWV using the two daily measurements from 2012 to 2013. We find the mean annual precipitable water measured at LDT from 2012 to 2013 is $9.5\pm5.2$~mm with the lowest monthly mean values at just under 5~mm. We also find that over 10 years from 2010 to 2019, the LDT annual mean value only varies by 1.5\%.

Due to the high annual PWV at the LDT site, we expect the near-infrared telluric bands to often be deep and near saturation. We also find that the width and depth of the bands change frequently throughout the duration of an observing night. As a result, standard reduction methods like those used in Spextool are not always ideal for telluric correction. 

\begin{figure}[!t]
    \centering
    \includegraphics[width=\linewidth]{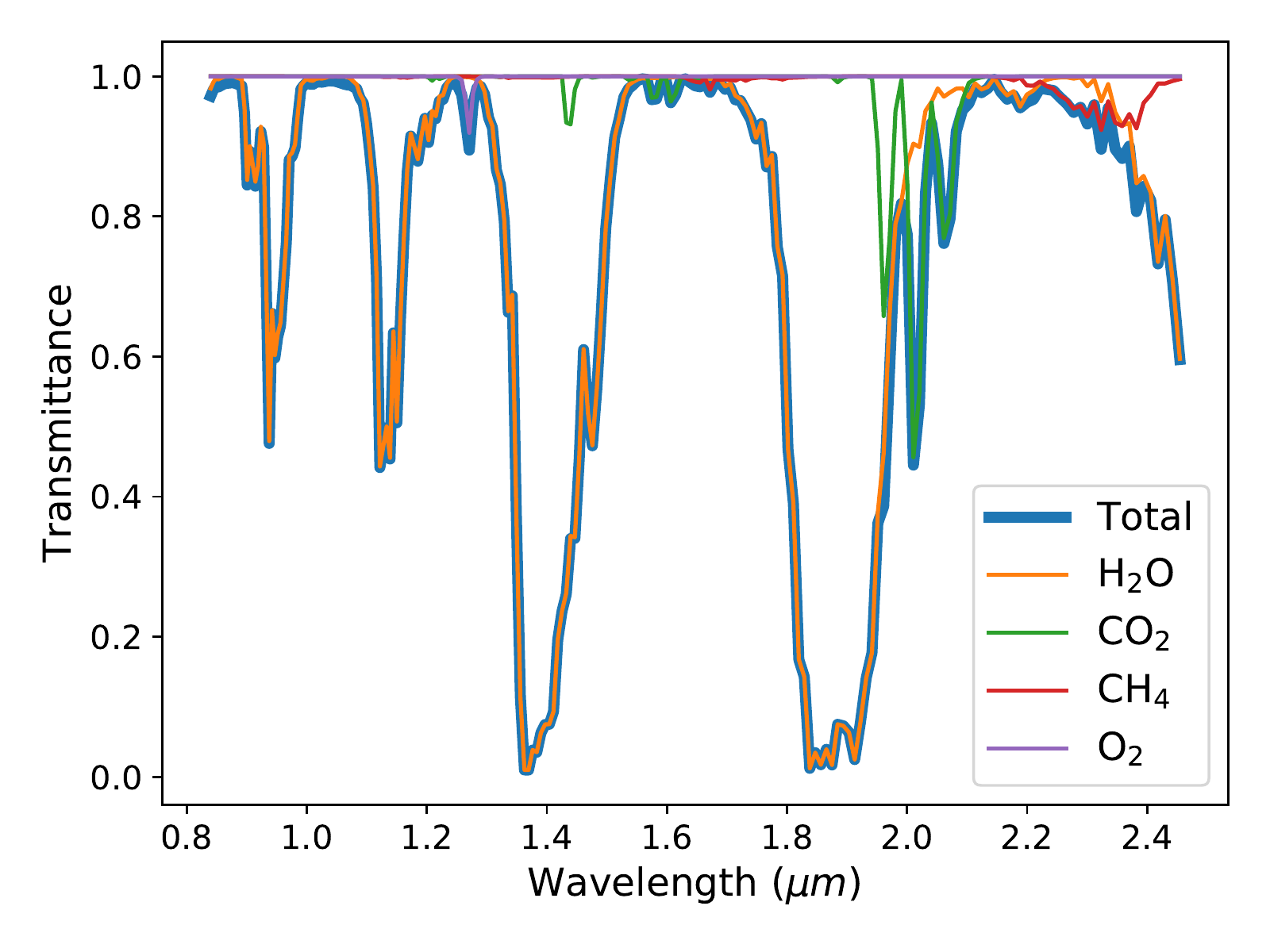} 
    \caption{Atmospheric transmission spectrum generated from the Planetary Spectrum Generator tool at UT 02:50 2018 September 17 at an airmass of 1.02 from the LDT using the Planetary Spectrum Generator. Major contributing species to the total atmospheric transmission are indicated. 
    \label{fig:atm_trans}}
\end{figure}

We have implemented use of the PSG \citep{2018Villanueva} for small body data reduction to improve telluric removal in NIHTS data when the PWV is high. We outline here the reduction of asteroid spectra, though many of these steps could be generalized to other science cases. An example of the improvement of telluric corrections using PSG is shown in red in Figure~\ref{fig:Spextool_vs_PSG}. PSG uses data from the Modern-Era Retrospective Analysis for Research and Applications (MERRA-2) satellite database \citep{2017Gelaro} to extract atmospheric profiles and molecular abundances at a cadence of 180 minutes and $\sim$1~km spatial resolution \citep{2018Villanueva}. This provides a means to correct for atmospheric telluric features which depend strongly on time and location of the observation.  

PSG has both an online tool\footnote{https://psg.gsfc.nasa.gov} and an API. We configure the inputs for our observing conditions (e.g., viewing geometry, UT date and time, sub-observer longitude and latitude) and generate an atmospheric transmission profile (Figure~\ref{fig:atm_trans}).

\begin{figure}[!t]
    \centering
    \includegraphics[width=\linewidth]{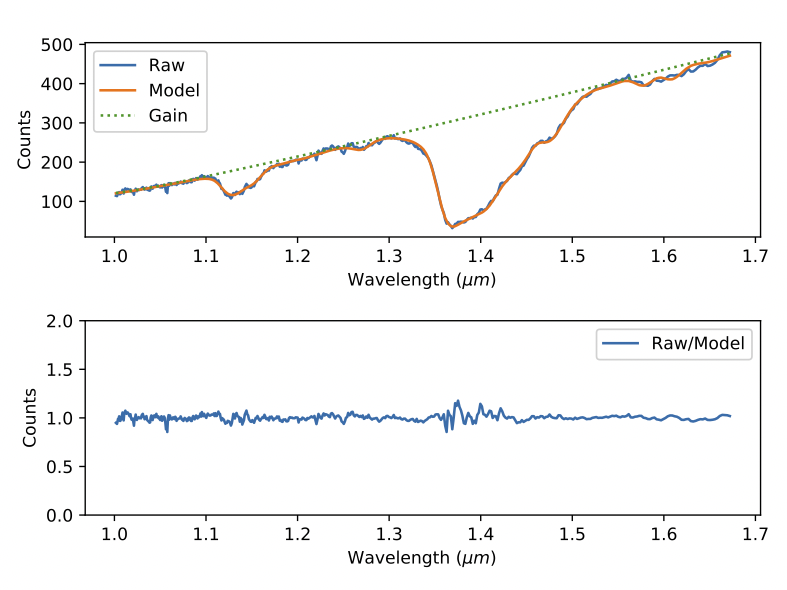} 
    \caption{Model fit to raw NIHTS data using the atmospheric transmission profile derived with PSG. The top panel shows the reduced data (blue) and the best fit model (orange), which is the combination of atmospheric absorption plus a continuum or gain profile (green). The data are cropped from 1.0--1.7~microns to optimize the wavelength region around 1.5~microns where we were searching for water-ice absorption. The bottom panel shows the residual of the model and the reduced standard star data. The corresponding PWV for these data is 5.8~mm.
    \label{fig:PSG}}
\end{figure}

Using a 1~D flattened spectrum of our data, we find a best fit to the atmospheric transmission profile using a stellar continuum and scaling of the atmospheric molecular abundances. In Figure~\ref{fig:PSG}, we show the best fit model to observations of a G2~V standard star from 1.0 to 1.7~microns. The longer wavelengths were saturated due to high thermal background and are therefore excluded in this figure. 
The output of PSG yields a file with values for wavelength, input data, model, and continuum (referred to as gain).

To obtain our final reduced spectrum, we remove the modeled atmospheric absorption features from both the asteroid and the standard star, and divide by the spectrum of the standard star (e.g., A0~V, G2~V). As demonstrated in Figure~\ref{fig:PSG}, we find great improvement to the telluric absorption bands using this technique.


\section{Science Cases} \label{sec:science}
We outline here select science use cases for NIHTS that are currently being pursued by LDT observers: Kuiper Belt Objects (KBOs), asteroids, comets, low mass stars, and exoplanet host stars. More detail is provided in the examples for minor planets in the solar system as these represent the predominant objects of study for the instrument. Other science use cases not currently being pursued could include, but are not limited to, compositions of planetary nebulae, searching for and characterizing nearby AGN, and detecting and mapping  emission lines from galaxies at high redshifts to yield information about their star formation histories.


\subsection{Kuiper Belt Objects}

NIHTS was originally designed as a tool to map the spectral diversity of KBOs, thus enabling investigations into the relationship between chemistry, collisions, and dynamics in the outer solar system. 

Near-infrared reflectance spectroscopy is well suited to provide compositional information of KBOs. The near-infrared wavelength region offers the best compromise between atmospheric transparency, solar flux, and the existence of diagnostic spectral features. Many cosmochemically plausible materials, such as molecular ices, simple organic compounds, and silicate minerals can be identified from their characteristic infrared vibrational absorption bands. We already know from existing spectroscopic and broad-band color data that the Kuiper Belt population is compositionally diverse (e.g., \citealt{2007Cruikshank, 2006Delsanti, 2008Tegler, 2012Fraser}). Since reflectance studies only probe the surfaces of KBOs, surface-altering processes such as photolysis and micrometeorite bombardment can strongly influence what can be detected (e.g., \citealt{2008Hudson, 2006Bennett, 2010Delsanti}), and the effects of these factors must be controlled by means of statistical comparison. 

Spectra of the largest and brightest KBOs (e.g., Pluto, Eris, Sedna, Quaoar, Haumea, Makemake) have been readily collected, but unfortunately, only a small, poorly-representative sample of  midsize KBOs (500--1000~km) have been explored spectroscopically (e.g., \citealt{2008Barkume, 2009Guilbert, 2011Barucci, 2012Brown}), owing to the significant time required on large-aperture facilities.

Smaller KBOs ($H>$6) are largely unexplored, except by using Centaurs and comets as proxies, which may not be appropriate given their likely alteration by proximity to the sun and their distinct albedo and color properties \citep{2003Tegler+Romanishin, 2003Tegler, 2008Barkume, 2012Brown}. Some dynamical classes of KBOs are under-sampled, such as the dynamically ``cold" classical population, which probably includes the least altered remnants of the outer reaches of our protoplanetary disk, as evidenced by their dynamically unexcited orbits, their high rates of binarity, and their  distinct color and albedo properties (e.g., \citealt{2002Trujillo+Brown, 2005Grundy, 2006Gulbis, 2008Noll, 2019Thiroin}). 
\begin{figure}[!t]
    \centering 
    \includegraphics[width=\linewidth]{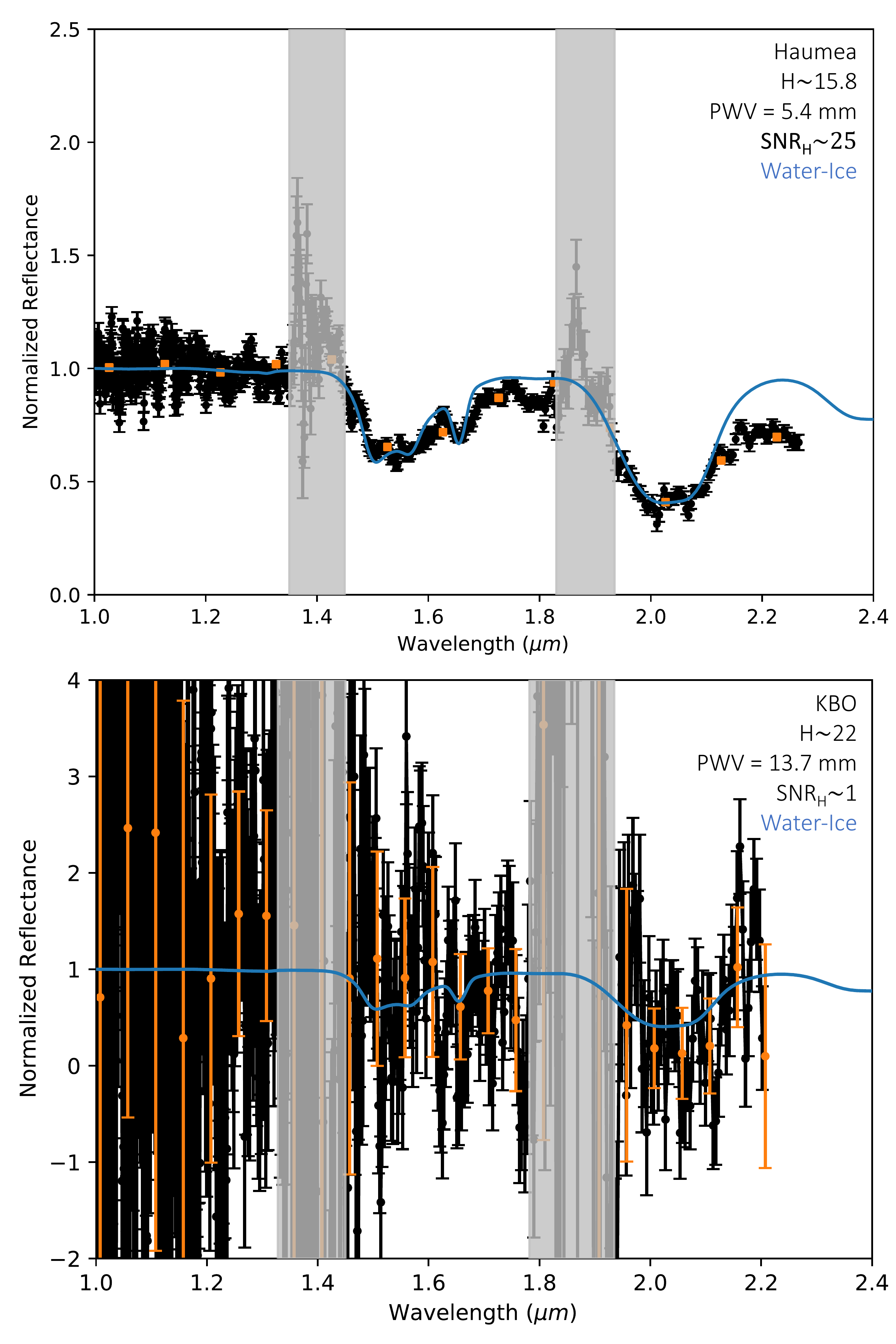}
    \caption{In the top panel we have a NIHTS spectrum of Haumea shown in black with 1$\sigma$ uncertainties. Haumea is known to be dominated by surface water-ice and reveals deep water-ice absorption features in the spectra \citep{2007Trujillo}. In the bottom panel, we have a small KBO ($V$=20.44; $H\sim$22). The total integration time is 8 minutes for S/N=1. The data shown in orange are binned into 0.01~micron wavelength bins. Overplotted in blue is a model spectrum of water-ice with 80~micron grains \citep{2000Buie} showing clear absorption features at 1.5, 1.65, and 2~microns, consistent with those for Haumea. The gray vertical bars represent the wavelength regions where telluric absorption features are strong.}
    \label{fig:KBO}
\end{figure}

Large KBOs show spectroscopic evidence of water-ice on their surface, whereas small KBOs do not. There is visible evidence of this transition in objects between absolute magnitudes of $\sim$3--5 in the 2~micron water absorption feature \citep{2011Barucci, 2012Brown}. Currently, this transition between water-rich and water-poor surfaces is not clearly understood. These objects are distant and typically faint ($V\ge$20), and therefore not easily detected in the near-infrared. However, the combination of LMI, NIHTS, and LDT tracking will make observations of these objects more feasible. NIHTS will allow for the study of water-ice abundances near the transition size. In Figure~\ref{fig:KBO} top panel, we show a NIHTS spectrum of (136108) Haumea ($H_{abs}$=0.2; $V$=17.3), dominated by water-ice with strong absorption features visible at 1.5, 1.65, and 2~microns. In the bottom panel, we show the NIHTS spectrum of a small KBO ($H_{abs}$=9; $V$=20.44) with a S/N=1 and total integration time of 8 minutes. The size of this object suggests that the surface may be poor, however the spectrum does not rule out a water-ice absorption feature at 2 microns. 


\subsection{Asteroids} \label{subsec:Ast}

Over the past 20 years, near-infrared spectroscopy has been a primary tool for characterizing the key physical properties (e.g., grain size, mineralogy, albedo, extent of space weathering) of Main Belt and near-Earth asteroids (e.g., \citealt{2019Binzel}). Asteroids are relatively unchanged remnant debris from early solar system formation, so understanding their physical properties allows for better constraints on formation and evolution models of our solar system. 

NIHTS is well suited to continuing these efforts, especially for faint objects (15.5$\le H\le$19.5) easily detected by visible surveys, but not easily observed with spectroscopic characterization. NIHTS is also well suited for small near-Earth objects (50~m$\le D\le$1~km) that get bright when passing close to the Earth but have short observing windows. Important for near-Earth asteroid spectroscopy is the ability to accurately track the telescope at high non-sidereal rates. Lowell has a long history of supporting telescopic capabilities necessary to observe solar system targets; these non-sidereal capabilities are fully implemented at LDT.

\begin{figure}[!t]
    \centering 
    \includegraphics[width=\linewidth]{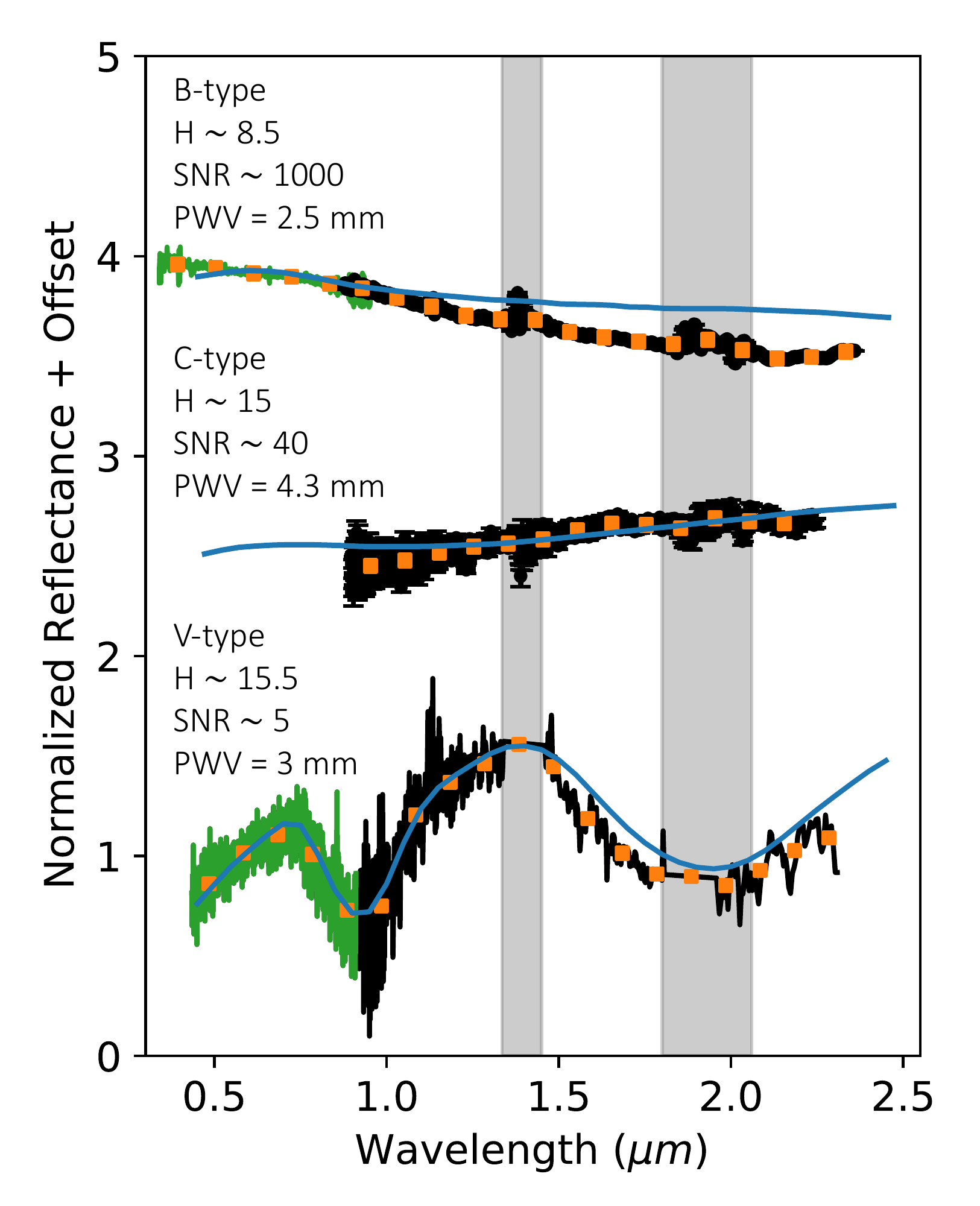}
    \caption{Asteroid taxonomic types observed with DeVeny (green) and NIHTS (black) with 1$\sigma$ error bars. The data shown in orange are binned into 0.01~micron wavelength bins. Overplotted in blue are the Bus-DeMeo mean spectra for the B-, C-, and V-type taxonomies \citep{2009DeMeo}. The B-type asteroid shown here is Phaethon (A.\ Gustafsson et al. 2021, in preparation). The vertical gray bars represent regions dominated by telluric absorption. These reductions were performed using Spextool.}
    \label{fig:Ast_Spec}
\end{figure}

The setup at LDT with a visible spectrograph (a single-order, long slit instrument called the DeVeny Spectrograph; \citealt{2014Bida}), NIHTS, and LMI all simultaneously co-mounted on the instrument cube allows for comprehensive and efficient characterization of asteroids, particularly near-Earth asteroids that may have short or limited observing windows. Near-infrared spectroscopy with NIHTS can be collected simultaneously with LMI images, which can provide broad-band photometric colors in select visible bands, rotational lightcurve information, and astrometry for orbit refinement. The switch between instruments from NIHTS to DeVeny, achieved by retracting the NIHTS dichroic and extending a fold mirror to feed the DeVeny side port, takes less than 30~seconds. While the target will need to be reacquired with each switch, the overhead remains low and allows for efficient collection of a full reflectance spectrum from 0.35 to 2.45~microns (e.g., Figure~\ref{fig:Ast_Spec} B-type and V-type). 

In Figure~\ref{fig:Ast_Spec}, we show three different asteroid taxonomic types \citep{2009DeMeo} observed with NIHTS ranging from spectrally featureless primitive compositions (B/C-types) to a heavily altered achondritic composition (V-type) with prominent 1 and 2~micron absorption features. In the following we highlight two specific objects, near-Earth asteroid 3200 Phaethon (B-type) and Main Belt asteroid 93 Minerva (C-type), to demonstrate the unique capability of NIHTS and the other LDT instruments to build comprehensive and complex data sets within a single night.

The unique active asteroid Phaethon, dynamically related to the parent body of the Geminid meteor shower \citep{1983Whipple}, made a close approach to Earth in 2017 December. The target passed within 0.07~AU and provided a rare opportunity to search for rotationally resolved compositional differences across the surface of the $\sim$5~km body. We observed Phaethon before and after closest approach between 2017 December 14 and 2017 December 18 obtaining both visible and near-infrared spectra with DeVeny and NIHTS, and simultaneous visible lightcurves with LMI through the dichroic in SDSS r$'$ and narrowband comet filters \citep{2020Ye}. Because the transition time between DeVeny and NIHTS is so fast, we were able to continually monitor Phaethon across the full wavelength range by switching instruments approximately every 10~minutes. 
An example of Phaethon's DeVeny+NIHTS spectrum is shown in Figure~\ref{fig:Ast_Spec} as the B-type object. The search for rotational heterogeneity in the NIHTS data is ongoing work that will be published in a future paper (A.\ Gustafsson et al. 2021, in preparation).

\begin{figure}[!t]
    \centering 
    \includegraphics[width=\linewidth]{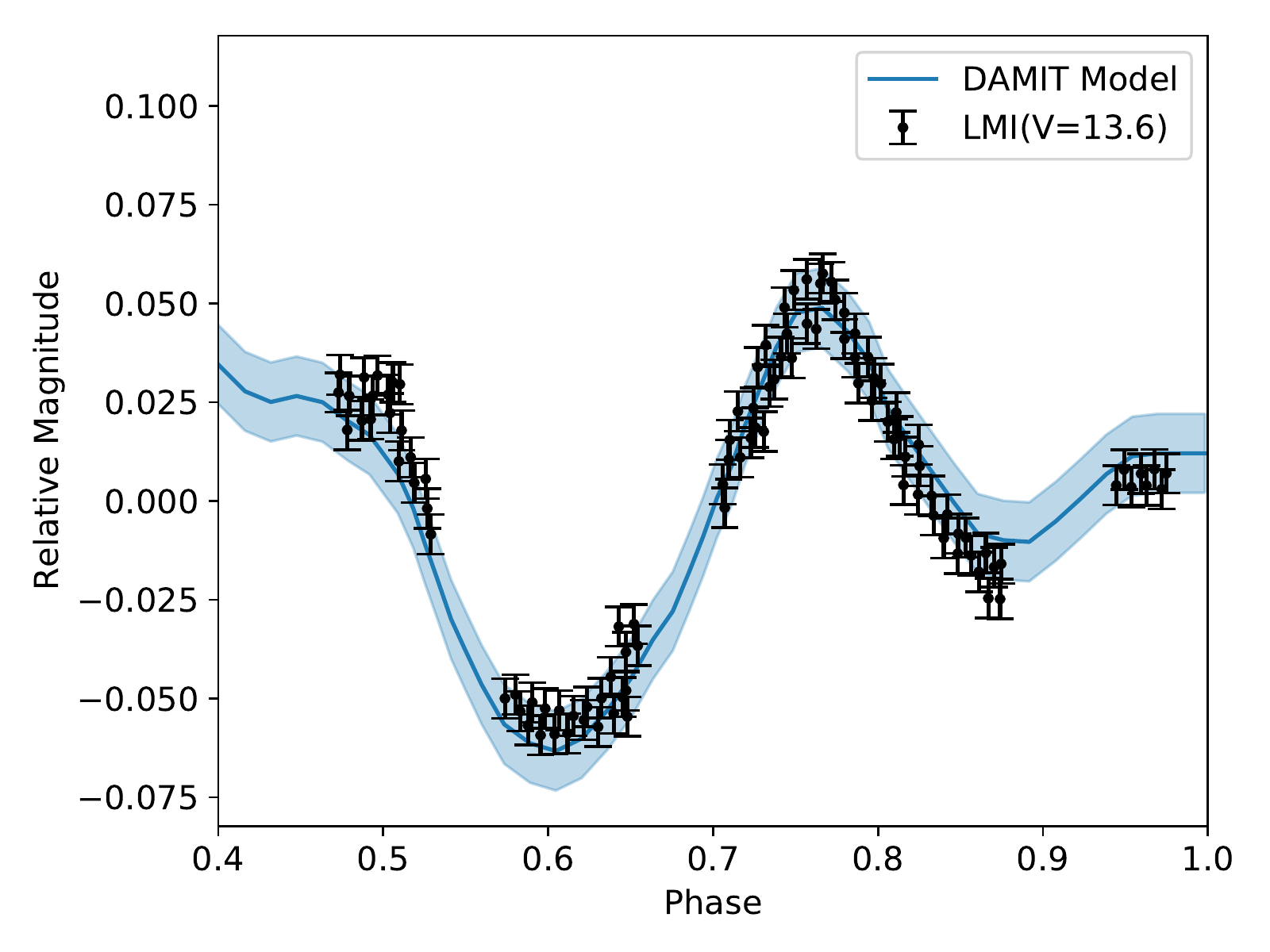}
    \caption{Minerva lightcurve from LMI observed simultaneously with NIHTS through the dichroic in SDSS r$'$. Overplotted in blue is the synthetic lightcurve derived from the Minerva shape model with a 0.01 magnitude uncertainty shown as the shaded region.}
    \label{fig:Minerva_LC}
\end{figure}

The simultaneity of NIHTS observations with LMI through the dichroic allows for rotational context when searching for surface heterogeneity. Main Belt asteroid 93 Minerva is a primitive triple asteroid system observed with NIHTS and LMI simultaneously in 2020 January. Figure~\ref{fig:Minerva_LC} shows the lightcurve obtained through the dichroic that will be used to provide rotational context for ongoing work to understand the spectral characteristics of the surface of Minerva. The photometry matches well with a synthetic lightcurve derived from the Minerva shape model in the DAMIT database \citep{2010Durech}. An example of a NIHTS spectrum of Minerva collected simultaneously with the LMI lightcurve is shown in Figure~\ref{fig:Spextool_vs_PSG}. A detailed analysis of the full suite of these observations is presented in McAdam \& Gustafsson (submitted).

\subsection{Comets}

\begin{figure*}[!t]
    \centering 
    \includegraphics[width=0.8\linewidth]{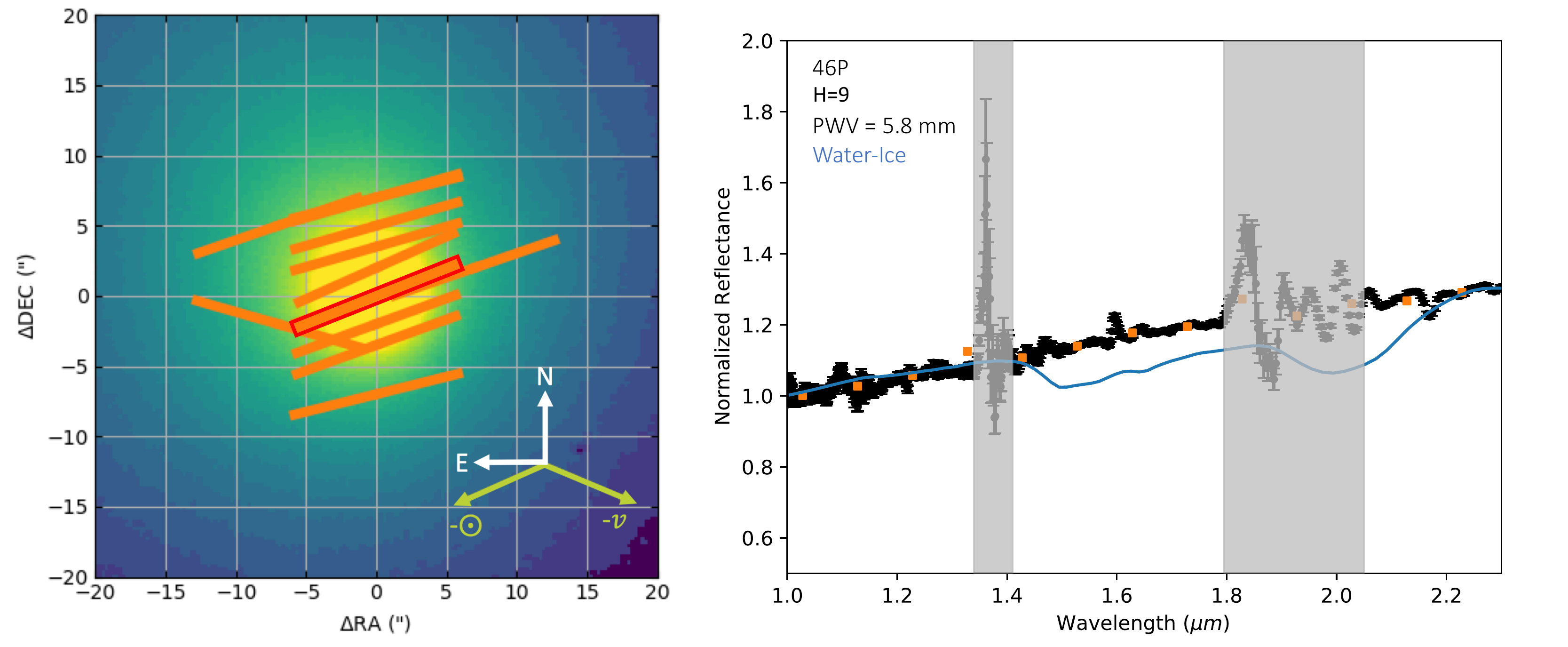}
    \caption{Left: LMI SDSS $r$~band image with the 1\farcsec34 NIHTS slitlet overplotted. Right: NIHTS spectrum of 46P/Wirtanen (from the slitlet outlined in red) shows no obvious water-ice absorption features at 1.5 or 2~microns. The data shown in orange are binned into 0.01~micron wavelength bins. The gray vertical bars represent the wavelength regions where telluric absorption features are strong. 46P/Wirtanen was observed at a $H$~band magnitude of $\sim$9. Overplotted in blue is a model spectrum of a coma with 10\% 1~micron sized water-ice particles and 90\% amorphous carbon (Provided by Bin Yang).}
    \label{fig:comet_spec}
\end{figure*} 

Comets are icy bodies believed to have formed beyond the snow line \citep{2012AHearn} and thus contain some of the most primitive extant material in our solar system \citep{2011AHearn}. The primary constituent of cometary nuclei is water ice, yet it is difficult to make direct observations of obscured cometary nuclei.  Furthermore, pristine material lies buried within the nucleus \citep{2004Bockelee, 1993Mumma} and might be different than that observed on the processed outer layer of comets. Most comets are small solar system bodies with elliptical orbits that bring them close to the Sun and then out beyond the solar system's water-ice line around 3 AU. When a comet makes a close passage to the Sun, volatiles are outgassed and water-ice grains can sometimes be detected in the coma (e.g., \cite{2009Yang, 2014Protopapa}). 

During cometary close approaches to the Earth, NIHTS has the ability to resolve water-ice characteristics both near the nucleus and along jet features of comets due to the realization of high spatial resolutions during such events. Simultaneous LMI$+$NIHTS observations in narrow-band comet filters (e.g., \emph{CN}, \emph{OH}; \citealt{2000Farnham}) are critical for determining real time morphology and NIHTS slit orientation relative to morphological features of the comet. The narrowband comet filter CN in particular is commonly used to track jet features which can be used to constrain the comet rotation period \cite[e.g.,][]{2019Schleicher}. The moderate field of view (4$'\times6'$) passed through the dichroic to LMI allows for real time decisions regarding slit position which are not easily accommodated at other telescopic facilities. Implementing similar techniques collected during close approach events will be crucial for matching morphology to the NIHTS slit orientation. In the following we provide an example for comet 46P/Wirtanen, which made a close approach to the Earth on UT 2019 December 16 coming within 0.077~au. This work will be presented in a future publication (A.\ Gustafsson et al. 2021, in preparation).

During the 46P flyby, a spectral data cube of observations was created with NIHTS observations by performing a slit scanning technique (i.e., stepping the slit across the comet after each exposure sequence) to obtain spatially resolved data centered on the nucleus (Figure~\ref{fig:comet_spec}, left). The spectral map of 46P is not a full coverage map, but probes radial and azimuthal profiles across the coma. Simultaneous LMI+NIHTS observations in narrow-band comet filters were utilized for determining real time coma morphology and slit placement with NIHTS. An example of one spectrum of 46P/Wirtanen is shown in the right panel of Figure~\ref{fig:comet_spec}. In this location in the coma, there is no obvious water-ice absorption feature at 1.5 or 2~microns, however, upper limits can be placed on the presence of water-ice using radiative transfer modeling \citep[e.g.~][]{2009Yang}.


\subsection{Low Mass Stars} \label{subsec:low_mass_stars}

\begin{figure}[!t]
    \centering 
    \includegraphics[width=\linewidth]{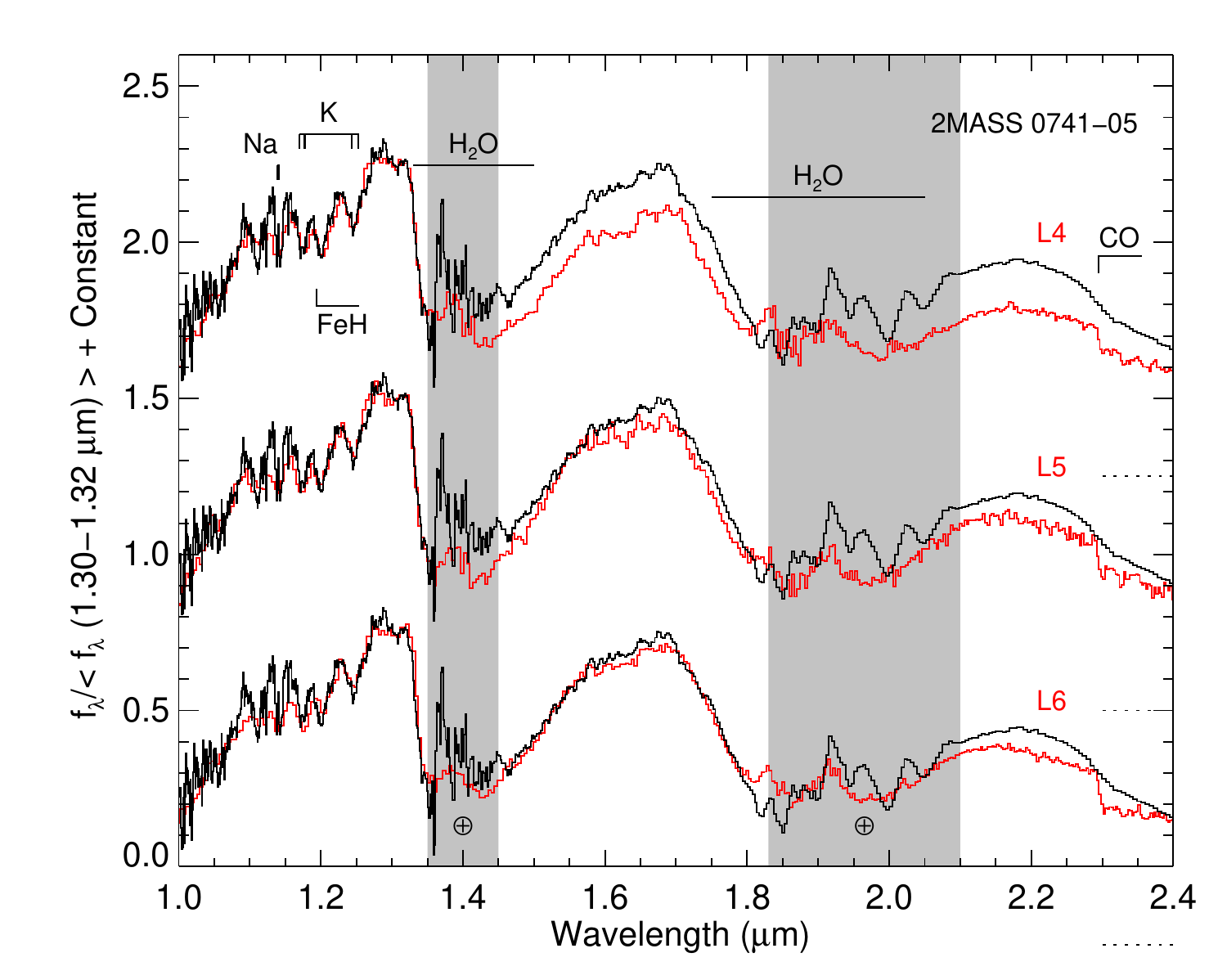}
    \caption{NIHTS spectrum of 2MASS 0741-05 (black) with comparison spectra of L dwarf spectral standards (red) from \cite{2010Kirkpatrick}. Prominent atomic and molecular features are labeled. Regions with strong telluric absorption features are indicated with gray bars. These reductions were performed using Spextool.}
    \label{fig:LowMass}
\end{figure}

The near-infrared spectra of low-mass stars and brown dwarfs are defined by broad molecular absorption features of TiO, VO, CrH, FeH, $H_2$O, CO, CH$_4$, and NH$_3$. Much of our understanding of these cool objects has come from the study of low spectral resolution, broad-wavelength spectra (e.g., \citealt{2008Cushing, 2009Stephens}) for stars with $H$=11.5-14 and NIHTS ability to push toward $H$=19. The wavelength coverage of NIHTS is particularly well suited to the study the late-type M, L, and T dwarfs because their SEDs peak at these wavelengths. For example, \cite{2018Cushing} used NIHTS to confirm that the candidate L dwarf 2MASS J07414279-0506464 was indeed a mid-type L dwarf with a spectral type of L5 (Figure~\ref{fig:LowMass}). \cite{2018Cushing} collected 10$\times$60~s exposures in the 1$\farcsec$34 NIHTS slit of 2MASS J07414279-0506464 which has a $J$~band magnitude of $14.173\pm0.050$ \citep{2018Scholz}.

The relative efficiency with which a low-resolution near-infrared spectrum can be obtained with NIHTS makes it an ideal instrument for efficient followup observations of candidate  M, L, and T dwarfs identified in photometric or proper motion searches (e.g., \citealt{2016Schneider, 2016Kirkpatrick, 2014Luhman}).

Stellar flares may play an important role in the main sequence evolution of low mass stars \citep{2012Davenport}. Recent studies have focused on correlating M dwarf flare rates with mass and metallicity to constrain ages of stellar populations (e.g., \citealt{2008West, 2009Kowalski, 2010Hilton}). Visible studies by \cite{2009Kowalski} using SDSS find the flare rates to decrease with peak flare luminosity and increase with stellar spectral type.

Few studies of M dwarf flares have been conducted in the near-infrared (e.g., \citealt{1988Rodono, 1995Panagi, 2012Davenport}). \cite{2012Davenport} performed the first statistical characterization of M dwarf flares in the near-infrared for the magnitude range 10$\le$H$\le$16. Flares in the \emph{J}-band are predicted to be detected most frequently in spectral types from M4 to M6 with amplitudes of $\Delta J\ge$0.01~mag \citep{2012Davenport}. NIHTS is well suited to better characterize M4--M6 flare rates in the near-infrared at these magnitudes and fainter.

\subsection{Exoplanet Transits} \label{subsec:exoplanets}

The atmosphere of an exoplanet provides clues to that planet's formation and evolutionary history, current climate, and habitability \citep{2018Kreidberg}. Spectroscopy of exoplanets in transit can inform the vertical profile of the atmosphere, making it possible to determine physical parameters connected to scale heights of various atmospheric constituents. In ideal cases, the amplitude of spectral features from a transiting exoplanet is $\sim$0.1\% making identification extremely difficult. For Earth like exoplanets, the amplitude can be two to three orders of magnitude smaller \citep{2018Kreidberg}. The technique of using multicolor observations of a planetary transit to better constrain physical characteristics of the planet was first demonstrated by \cite{2000Jha} with low-resolution spectrophotometric lightcurves using broadband filters. 

Characterization of exoplanetary atmospheres is challenging from the ground, particularly at near-infrared wavelengths due to the variable transparency of Earth's atmosphere and the rapidly changing line-of-sight properties (e.g., airmass, precipitable water, cloud cover) that can influence spectral data \citep{2018Kreidberg}. Simultaneously observing comparison field stars can help to mitigate these effects. For photometric observations, the comparison stars need to be in the same field of view, while for spectroscopy the science target and comparison star would need to fall on the same slit mask. Differential spectroscopy of exoplanet transits has been measured at visible wavelengths using a multi-object slit mask to measure the exoplanet and nearby reference stars of similar brightness \citep{2010Bean, 2011Bean, 2013Gibson, 2014Stevenson, 2017Huitson}. This technique, first developed by \cite{2010Bean} avoids time-dependent slit losses due to variable seeing and can reach spectrophotometric precision of $\sim10^{-4}$.

This technique has been implemented with the near-infrared spectrograph SpeX and visible imager MORIS on the NASA IRTF with a measured transit depth of $\sim$0.03-0.09\% \citep{2014Schlawin}. Similarly, the unique design of NIHTS also allows for simultaneous observations of exoplanet transits in the near-infrared utilizing the two 4$''\times$12$''$ SED slitlets, whose centers are separated by 84$''$, to perform differential time-series spectrophotometry with verification of the exoplanet transit from the visible LMI lightcurve. This is a science use case that is actively being analyzed by LDT partners who collected data with both SED slits of a transiting planet and a reference star. A simultaneous LMI lightcurve was also collected and a transit is confirmed. The is ongoing work that will be published in a separate paper.



\section{Conclusion}

We have presented results from the early performance characteristic and commissioning of NIHTS on the 4.3~m Lowell Discovery Telescope. NIHTS is a low-resolution prism spectrograph with no moving parts and 7 different slitlet widths from 0\farcsec27--4$''$ wide. NIHTS is a compact instrument with high efficiency and high throughput ($\sim$40\%). We have presented reduction procedures for NIHTS with two approaches to telluric correction, all of which are publicly available. Finally, we presented NIHTS spectra from science cases connected to small bodies in the solar system, low mass stars, and exoplanet transits, topics that are being actively pursued by Lowell Observatory, Northern Arizona University, and Lowell Discovery Telescope partners.


\acknowledgements

Funds for the construction of NIHTS were obtained from NASA's Planetary Astronomy and Planetary Major Equipment programs (PI: Henry Roe). A grant awarded to the Mission Accessible Near-Earth Objects Survey (MANOS, PI: N.\ Moskovitz), funded by the NASA Near-Earth Object Observations Program \#NNX17AH06G, has provided important support for finishing many of the commissioning tasks. The Mt.\ Cuba Foundation funded the NIHTS dichroic coatings for which simultaneous observing with NIHTS and LMI is now possible. We are grateful to the excellent staff at LDT who who helped to support the nightly operations and commissioning of NIHTS.

The overall system design and optical design of NIHTS was performed by Ted Dunham. Ted Dunham and Rich Oliver did the electrical design and construction of the instrument. Mechanical design was done by James French, Tomas Chylek, Frank Cornelius, Tom Bida, and Ted Dunham. Most of the NIHTS instrument parts were made by Steve Lauman and Jeff Gehring or by Boston University. The control software and user interface for the NIHTS ZTV was written by Henry Roe, while the NIHTS LOUI was written by the late Peter Collins, Saeid Zoonemat Kermani, and Dyer Lytle. 

These results made use of the Lowell Discovery Telescope at Lowell Observatory. Lowell is a private, non-profit institution dedicated to astrophysical research and public appreciation of astronomy and operates the LDT in partnership with Boston University, the University of Maryland, the University of Toledo, Northern Arizona University and Yale University. The Large Monolithic Imager was built by Lowell Observatory using funds provided by the National Science Foundation (AST-1005313).

The authors thank Maggie McAdam and Andrew McNeill for supplying LMI photometry and the Minerva shape model to include in this work. We also thank Geronimo Villanueva and Silvia Protopapa for assisting in deriving the procedure to incorporate PSG into NIHTS reductions for improved telluric corrections. Lastly, the authors thank Bin Yang for supplying a spectrum of water-ice spectrum to compare with comet 46P for this work and Joe Llama for guidance and feedback writing the sections on exoplanet transits and M dwarf flares. 

\bibliography{mybibliography}

\end{document}